\documentclass[trackchanges,twocolumn,12pt,resetfootnote]{aastex701}

\usepackage{hyperref}				% To use hypertextual links (e.g. links to websites). Per usare riferimenti ipertestuali (e.g. link a siti web).
\hypersetup{colorlinks, citecolor=blue,						% Chose the color for citations. Colore dei link per le citazioni.
 linkcolor=black						% Color of internal links (change box color with linkbordercolor). Scelgo il colore per i links.
 }

\newcolumntype{L}[1]{>{\raggedright\let\newline\\\arraybackslash\hspace{0pt}}m{#1}}
\newcolumntype{C}[1]{>{\centering\let\newline\\\arraybackslash\hspace{0pt}}m{#1}}
\newcolumntype{R}[1]{>{\raggedleft\let\newline\\\arraybackslash\hspace{0pt}}m{#1}}

\newcommand{\A}{\text{\normalfont\AA}}

\def \h2{{\rm H_{2}}}

\def \dn4000{D_{{\rm n}}(4000) }

\def \x{$\times$}

\usepackage[section]{placeins}

\usepackage{array}
\usepackage{amsmath,amssymb}
\usepackage[flushleft]{threeparttable}
\usepackage{array,booktabs,makecell}
\usepackage[font=small]{caption}

\begin{document}

\title{CHAMPS: Data Reduction and Detection of $\mathbf{\sim3300}$ Dusty Galaxies at 1.2\,mm}

% \author{Felix Martinez III and CHAMPS Team}
% \affiliation{Rochester Institute of Technology}
% \email[hide]{felixmartz123@gmail.com}  

\suppressAffiliations

%%% First Author %%%
% Felix
\author[0000-0002-9883-1413]{Felix Martinez III}
\email{fm5957@rit.edu}
\affiliation{Laboratory for Multiwavelength Astrophysics, School of Physics and Astronomy, Rochester Institute of Technology, 84 Lomb Memorial Drive, Rochester, NY 14623, USA}
\correspondingauthor{Felix Martinez III}

%%% First Tier %%%
% Core Data Reduction Team + CHAMPS PIs
% Jeyhan Karaltepe
\author[0000-0001-9187-3605]{Jeyhan S. Kartaltepe}
\affiliation{Laboratory for Multiwavelength Astrophysics, School of Physics and Astronomy, Rochester Institute of Technology, 84 Lomb Memorial Drive, Rochester, NY 14623, USA}
\email{jeyhan@astro.rit.edu}

% Andreas Faisst
\author[0000-0002-9382-9832]{Andreas L. Faisst}
\email{afaisst@caltech.edu}
\affiliation{Caltech/IPAC, MS 314-6, 1200 E. California Blvd. Pasadena, CA 91125, USA}

% Arianna Long
\author[0000-0002-7530-8857]{Arianna S. Long}
\affiliation{Department of Astronomy, The University of Washington, Seattle, WA 98195, USA}
\email{aslong@uw.edu}

% Jorge Zavala
\author[0000-0002-7051-1100]{Jorge A. Zavala}
\email{jzavala@umass.edu}
\affiliation{University of Massachusetts Amherst, 
710 North Pleasant Street, Amherst, MA 01003-9305, USA}

% Caitlin Casey
\author[0000-0002-0930-6466]{Caitlin M. Casey}
\email{cmcasey@ucsb.edu}
\affiliation{Department of Physics, University of California, Santa Barbara, Santa Barbara, CA 93106, USA}
\affiliation{Cosmic Dawn Center (DAWN), Denmark}

% Jed McKinney
\author[0000-0002-6149-8178]{Jed McKinney}
\altaffiliation{NASA Hubble Fellow}
\affiliation{Department of Astronomy, The University of Texas at Austin, 2515
Speedway Blvd Stop C1400, Austin, TX 78712, USA}
\affiliation{Cosmic Frontier Center, The University of Texas at Austin, Austin, TX 78712}
\email{}

% Ezequiel Treister
\author[0000-0001-7568-6412]{Ezequiel Treister}
\email{etreiste@astro.puc.cl}
\affiliation{
Instituto de Alta Investigaci{\'{o}}n, Universidad de Tarapac{\'{a}}, Casilla 7D, Arica, Chile}

% Maximilien Franco
\author[0000-0002-3560-8599]{Maximilien Franco}
\email{maximilien.franco@iap.fr}
\affiliation{Institut d’Astrophysique de Paris, UMR 7095, CNRS, Sorbonne Université, 98 bis boulevard Arago, 75014 Paris, France}

% Sune Toft
\author[0000-0003-3631-7176]{Sune Toft}
\affiliation{Cosmic Dawn Center (DAWN), Denmark} 
\affiliation{Niels Bohr Institute, University of Copenhagen, Jagtvej 128, DK-2200, Copenhagen, Denmark}
\email{sune@nbi.ku.dk}

% Manuel Aravena
\author[0000-0002-6290-3198]{Manuel Aravena}
\affiliation{Instituto de Estudios Astrofisicos, Facultad de Ingenieria y Ciencias, Universidad Diego Portales, Av. Ej\'ercito 441, Santiago 8370191, Chile}
\affiliation{Millenium Nucleus for Galaxies (MINGAL), Av. Ej\'ercito 441, Santiago 8370191, Chile}
\email{}

% John Silverman
\author[0000-0002-0000-6977]{John D. Silverman}
\affiliation{Kavli Institute for the Physics and Mathematics of the Universe (WPI), The University of Tokyo, Kashiwa, Chiba 277-8583, Japan}
\affiliation{Department of Astronomy, School of Science, The University of Tokyo, 7-3-1 Hongo, Bunkyo, Tokyo 113-0033, Japan}
\email{}

%%% Second Tier %%%
% Others who have actively contributed but are not in first tier including
% Hollis Akins
\author[0000-0003-3596-8794]{Hollis B. Akins}
\email{hollis.akins@gmail.com}
\altaffiliation{NSF Graduate Research Fellow}
\affiliation{The University of Texas at Austin, 2515 Speedway Blvd Stop C1400, Austin, TX 78712, USA}

% Hiddo Algera
\author[0000-0002-4205-9567]{Hiddo S. B. Algera}
\email{hsbalgera@asiaa.sinica.edu.tw}
\affiliation{ Institute of Astronomy and Astrophysics, Academia Sinica, 11F of Astronomy-Mathematics Building, No.1, Sec. 4, Roosevelt Rd,
Taipei 106319, Taiwan, R.O.C.}

% Jacqulen Antwidanso
\author[0000-0002-0243-6575]{Jacqueline Antwi-Danso}
\affiliation{University of Toronto, 50 St. George Street, Toronto, Ontario, Canada M5S 3H4}
\email{}

% Andrew Battisti
\author[0000-0003-4569-2285]{Andrew J. Battisti}
\affil{International Centre for Radio Astronomy Research, University of Western Australia, 35 Stirling Hwy, Crawley, WA 6009, Australia}
\affil{Research School of Astronomy and Astrophysics, Australian National University, Cotter Road, Weston Creek, ACT 2611, Australia}
\email{}

% Yingjie Cheng
\author[0000-0001-8551-071X]{Yingjie Cheng}
\email{yingjiec@uw.edu}
\affiliation{Department of Astronomy, University of Washington, Seattle, WA 98195, USA}
\affiliation{
University of Massachusetts Amherst, 710 North Pleasant Street, Amherst, MA 01003-9305, USA}

% Olivia Cooper
\author[0000-0003-3881-1397]{Olivia R. Cooper}
\email{ocooper@utexas.edu}
\affiliation{Department for Astrophysical \& Planetary Science, University of Colorado, Boulder, CO 80309, USA}

% Nicole Drakos
\author[0000-0003-4761-2197]{Nicole E. Drakos}
\email{ndrakos@hawaii.edu}
\affiliation{Department of Physics and Astronomy, University of Hawaii, Hilo, 200 W Kawili St, Hilo, HI 96720, USA}

% Carter Flayhart
\author[0009-0000-0272-5468]{Carter Flayhart}
\email{cbf2388@rit.edu}
\affiliation{Laboratory for Multiwavelength Astrophysics, School of Physics and Astronomy, Rochester Institute of Technology, 84 Lomb Memorial Drive, Rochester, NY 14623, USA}

% Fabrizio Gentile
\author[0000-0002-8008-9871]{Fabrizio Gentile}
\affiliation{CEA, Université Paris-Saclay, Université Paris Cité, CNRS, AIM, 91191, Gif-sur-Yvette, France}
\affiliation{INAF- Osservatorio di Astrofisica e Scienza dello Spazio, Via Gobetti 93/3, I-40129, Bologna, Italy}
\email{}

% % Olivier Gilbert (whoops!)
% \author[0009-0004-2538-7237]{Olivier Gilbert}
% \email{ogilbert@umich.edu}
% \affiliation{University of Michigan: Ann Arbor, Michigan, US}

% Steven Gillman
\author[0000-0001-9885-4589]{Steven Gillman}
\affiliation{Cosmic Dawn Center (DAWN), Denmark}
\affiliation{DTU-Space, Technical University of Denmark, Elektrovej 327, DK-2800 Kgs. Lyngby, Denmark}
\email{}

% Ghassem Gozaliasl
\author[0000-0002-0236-919X]{Ghassem Gozaliasl}
\email{ghassem.gozaliasl@gmail.com}
\affiliation{Department of Computer Science, Aalto University, P.O. Box 15400, FI-00076 Espoo, Finland}
\affiliation{Department of Physics, University of, P.O. Box 64, FI-00014 Helsinki, Finland}

% Santosh Harish
\author[0000-0003-0129-2079]{Santosh Harish}
\email{harish.santosh@gmail.com}
\affiliation{Space Telescope Science Institute, 3700 San Martin Dr., Baltimore, MD 21218, USA}
\affiliation{Laboratory for Multiwavelength Astrophysics, School of Physics and Astronomy, Rochester Institute of Technology, 84 Lomb Memorial Drive, Rochester, NY 14623, USA}

% Simon Hempel-Costello
\author[0009-0006-8917-119X]{Simon Hempel-Costello}
\email{shempelc@caltech.edu}
\affiliation{California Institute of Technology, 1200 E. California Blvd., Pasadena, CA, 91125 USA}

% Michaela Hirschmann
\author[0000-0002-3301-3321]{Michaela Hirschmann}
\affiliation{Institute of Physics, GalSpec, Ecole Polytechnique Federale de Lausanne, Observatoire de Sauverny, Chemin Pegasi 51, 1290 Versoix, Switzerland}
\affiliation{INAF, Astronomical Observatory of Trieste, Via Tiepolo 11, 34131 Trieste, Italy}
\email{}

% Olivier Ilbert
\author[0000-0002-7303-4397]{Olivier Ilbert}
\email{olivier.ilbert@lam.fr}
\affiliation{Aix Marseille Univ, CNRS, CNES, LAM, Marseille, France  }

% Bingcheng Jin
\author[0009-0003-4862-2925]{Bingcheng Jin}
\affiliation{University of Michigan: Ann Arbor, Michigan, US}
\email{bcjin@umich.edu}

% Shuowen Jin
\author[0000-0002-8412-7951]{Shuowen Jin}
\email{shuowen.jin@gmail.com}
\affiliation{Cosmic Dawn Center (DAWN), Denmark}
\affiliation{DTU Space, Technical University of Denmark, Elektrovej 327, 2800 Kgs. Lyngby, Denmark}

% Koki Kaiichi
\author[0000-0001-6874-1321]{Koki Kakiichi}
\affiliation{Cosmic Dawn Center (DAWN), Denmark}
\email{koki.kakiichi@nbi.ku.dk}

% Thresa Kelly
\author[0009-0005-9497-7289]{Thresa Kelly}
\email{tsk4215@rit.edu}
\affiliation{Laboratory for Multiwavelength Astrophysics, School of Physics and Astronomy, Rochester Institute of Technology, 84 Lomb Memorial Drive, Rochester, NY 14623, USA}

% Zihao Li
\author[0000-0001-5951-459X]{Zihao Li}
\affiliation{Cosmic Dawn Center (DAWN), Denmark}
\email{zihao.li@nbi.ku.dk}

% Daizhong Liu
\author[0000-0001-9773-7479]{Daizhong Liu}
\email{dzliu@pmo.ac.cn}
\affiliation{Purple Mountain Observatory, Chinese Academy of Sciences, 10 Yuanhua Road, Nanjing 210023, China}

% Lun-Jun Liu
\author[0009-0004-1270-2373]{Lun-Jun Liu}
\email{lliu@caltech.edu}
\affiliation{California Institute of Technology, 1200 E. California Blvd., Pasadena, CA, 91125 USA}

% Sinclaire Manning
\author[0000-0003-0415-0121]{Sinclaire M. Manning}
\email{smanning@astro.umass.edu}
\affiliation{Department of Astronomy, University of Massachusetts Amherst, 710 N Pleasant Street, Amherst, MA 01003, USA}

% Henry Joy McCracken
\author[0000-0002-9489-7765]{Henry Joy McCracken}
\email{hjmcc@iap.fr}
\affiliation{Institut d’Astrophysique de Paris, UMR 7095, CNRS, and Sorbonne Université, 98 bis boulevard Arago, F-75014 Paris, France}

% Romain Meyer
\author[0000-0001-5492-4522]{Romain A. Meyer}
\affiliation{Department of Astronomy, University of Geneva, Chemin Pegasi 51, 1290 Versoix, Switzerland}
\email{romain.meyer@unige.ch}

% Louise Paquereau
\author[0000-0003-2397-0360]{Louise Paquereau} 
\email{louise.paquereau@chalmers.se}
\affiliation{Department of Physics and Astronomy, Chalmers University of Technology, SE-412 96 Gothenburg, Sweden}

% Maria Pudoka
\author[0000-0003-4924-5941 ]{Maria Pudoka}
\email{}
\affiliation{Department of Astronomy and Steward Observatory
University of Arizona}

% Jason Rhodes
\author[0000-0002-4485-8549]{Jason Rhodes}
\affiliation{Jet Propulsion Laboratory, California Institute of Technology, 4800 Oak Grove Drive, Pasadena, CA 91001, USA}
\email{}

% Brant Robertson
\author[0000-0002-4271-0364]{Brant E. Robertson}
\affiliation{Department of Astronomy and Astrophysics, University of California, Santa Cruz, 1156 High Street, Santa Cruz, CA 95064, USA}
\email{brant@ucsc.edu}

% Giulia Rodighiero
\author[0000-0002-9415-2296]{Giulia Rodighiero}
\affiliation{Dipartimento di Fisica e Astronomia Galileo Galilei Universit{\`a} degli Studi di Padova, Vicolo dell’Osservatorio 3, 35122 Padova Italy}
\affiliation{Istituto Nazionale di Astrofisica (INAF), Osservatorio Astronomico di Padova,
Vicolo dell’Osservatorio 5, 35122, Padova, Italy}
\email{giulia.rodighiero@unipd.it}

% Rasha Samir
\author[0000-0003-2716-8332]{Rasha M. Samir}
\affiliation{Department of Astronomy, National Research Institute of Astronomy and Geophysics (NRIAG), Cairo, 11421, Egypt}
\email{rasha.samir@nriag.sci.eg}

% David Sanders
\author[0000-0002-1233-9998]{David B. Sanders}
\affiliation{Institute for Astronomy, University of Hawaii, 2680 Woodlawn Drive, Honolulu, HI 96822, USA}
\email{sandersd@hawaii.edu}

% Marko Shuntov
\author[0000-0002-7087-0701]{Marko Shuntov}
\email{marko.shuntov@nbi.ku.dk}
\affiliation{Cosmic Dawn Center (DAWN), Denmark} 
\affiliation{Niels Bohr Institute, University of Copenhagen, Jagtvej 128, DK-2200, Copenhagen, Denmark}
\affiliation{University of Geneva, 24 rue du Général-Dufour, 1211 Genève 4, Switzerland}

% Margherita Talia
\author[0000-0003-4352-2063]{Margherita Talia}
\email{margherita.talia2@unibo.it}
\affiliation{University of Bologna – Department of Physics and Astronomy ``Augusto Righi'' (DIFA), Via Gobetti 93/2, 40129 Bologna, Italy}
\affiliation{INAF-Osservatorio di Astrofisica e Scienza dello Spazio, Via Gobetti 93/3, 40129, Bologna, Italy}

% Mattia Vaccari
\author[0000-0002-6748-0577]{Mattia Vaccari}
\email{mattia.vaccari@uct.ac.za}
\affiliation{Inter-University Institute for Data Intensive Astronomy (IDIA),
Department of Astronomy, University of Cape Town, 7701 Rondebosch,
Cape Town, South Africa}
\affiliation{Department of Physics and Astronomy, University of the Western Cape,
7535 Bellville, Cape Town, South Africa}
\affiliation{INAF - Istituto di Radioastronomia, via Gobetti 101, 40129 Bologna, Italy}

% Francesco Valentino
\author[0000-0001-6477-4011]{Francesco Valentino}
\email{fmava@dtu.dk}
\affiliation{Cosmic Dawn Center (DAWN), Denmark} 
\affiliation{DTU Space, Technical University of Denmark, Elektrovej 327, 2800 Kgs. Lyngby, Denmark}

% Aswin Vijayan
\author[0000-0002-1905-4194]{Aswin P. Vijayan}
\affiliation{Astronomy Centre, University of Sussex, Falmer, Brighton BN1 9QH, UK} 
\email{}

% Feige Wang
\author[0000-0002-7633-431X]{Feige Wang}
\affiliation{University of Michigan: Ann Arbor, Michigan, US}
\email{fgwang@umich.edu}

% Wuji Wang
\author[0000-0002-7964-6749]{Wuji Wang}
\affiliation{IPAC, California Institute of Technology, 1200 E. California Blvd. Pasadena, CA 91125, USA}
\email{wujiwang@ipac.caltech.edu}

% Can Xu
\author[0000-0002-8437-6659]{Can Xu}
\affiliation{School of Astronomy and Space Science, Nanjing University, Nanjing, Jiangsu 210093, China}
\affiliation{Key Laboratory of Modern Astronomy and Astrophysics, Nanjing University, Ministry of Education, Nanjing 210093, China}
\affiliation{Kavli Institute for the Physics and Mathematics of the Universe (WPI), The University of Tokyo, Kashiwa, Chiba 277-8583, Japan}
\email{}

% Jinyi Yang
\author[0000-0001-5287-4242]{Jinyi Yang}
\email{}
\affiliation{University of Michigan: Ann Arbor, Michigan, US}
\affiliation{Department of Astronomy and Steward Observatory
University of Arizona}

% Yongda Zhu
\author[0000-0003-3307-7525]{Yongda Zhu}
\email{}
\affiliation{Department of Astronomy and Steward Observatory
University of Arizona}

% Siwei Zou
\author[0000-0002-3983-6484]{Siwei Zou}
\affiliation{Chinese Academy of Sciences South America Center for Astronomy, National Astronomical Observatories, CAS, Beijing 100101, China}
\affiliation{Departamento de Astronom\'ia, Universidad de Chile, Casilla 36-D, Santiago, Chile} 
\email{}

%%% Other collaborators???

\collaboration{all}{(Affiliations can be found after the references)}

\begin{abstract}

We present the reduction and source catalogs for the \textit{COSMOS High-$z$ ALMA-MIRI Population Survey} (CHAMPS), the largest Atacama Large (Sub)Millimeter Array (ALMA) blank field survey to date at 0.2\,deg$^2$. 
With observing frequency centered at $250\, $GHz ($1.2\,$mm) and bandwidth of 7.5 GHz, CHAMPS covers the JWST/NIRCam+MIRI footprint in the COSMOS field, taking advantage of the rich, multiwavelength ancillary datasets in COSMOS ranging from X-ray to radio.
The majority of the CHAMPS survey (90$\%$) has depths at or below 200 $\mu$Jy beam$^{-1}$ and with an average RMS of 146 $\mu$Jy beam$^{-1}$.
The reduced CHAMPS mosaic has a beam size of $\theta_\text{res} = 1.1\arcsec \times 1.1\arcsec$.
CHAMPS identifies sources via two detection methods, a blind search for high-SNR sources, and a prior-based search involving MIRI detections to identify faint sources.
The blind search yields a total of 595 sources down to an SNR $\ge4.5\sigma$ while the prior catalog yields a total of 3162 source down to an SNR $\ge3\sigma$, leading to a full sample of 3355 unique objects detected across both catalogs. 
We identify a unique population of 385 SNR $\ge5\sigma$ sources across the two catalogs; after excluding 9 likely spurious edge detections we define a robust science sample of 376 sources above $\ge5\sigma$.
% CHAMPS has identified 375 $5\sigma$ sources which have been verified through cross-matching ancillary datasets.
These sources have a median redshift of $\langle z_{1.2 \text{ mm}}\rangle = 2.54^{+0.89}_{-1.69}$ based on optical/near-IR/mid-IR photometric redshifts and high-quality spectroscopic redshifts, and hold a median stellar mass of $\langle\log (\rm M_\star/M_\odot)\rangle = 10.79^{+0.42}_{-3.08}$. 
% and hold a median stellar mass of $\langle\log(\rm M_\star/M_\odot)\rangle = 10.7$.
% We measure a median photometric redshift of $\langle z_\text{phot}\rangle = 2.54$ and a median stellar mass of $\langle\log(\rm M_\star/M_\odot)\rangle = 10.7$.
We have further identified 68/376 ($\sim18\%$) sources as candidate AGN through X-ray, UV-optical, infrared and radio selection techniques.
% , a fraction that matches previous estimates of AGN identified in sub-mm galaxies.

\end{abstract}

\section{Introduction} \label{sec:intro}

Detailed studies of dusty star-forming galaxies (DSFGs) at high redshift ($z>3$) utilizing sub-millimeter wavelengths have played an important role in understanding the rapid formation of early massive galaxies \citep{Blain_2002,Casey_2014,Hodge+da_Cunha_2020}.
DSFGs are characterized by infrared (IR) luminosities $L_{\rm{IR}} \gtrsim  10^{11} L_\odot$ and have intense dust-enshrouded star formation rates (SFRs) $\lesssim 100$\, M$_\odot \rm{yr}^{-1}$. 
At Cosmic Noon ($1.5<z<3$), the epoch defined by the peak in the cosmic star formation rate density (cSFRD) and supermassive black hole (SMBH) growth, 
% (related to active galactic nuclei, AGN),
these sources dominate star formation \citep{Madau_2014,Zavala_2021}.
DSFGs are believed to trace massive protocluster formation \citep{Casey2016}, and are the likely progenitors of the first generations of massive quiescent galaxies \citep{Toft_2014, Forrest_2020}. 
% However, their population density, and therefore their role in cosmic evolution, is severely under constrained at high-$z$ ($>$1-2 dex in variance at $z\gtrsim4$, \citealt{Long_2026}). 
Fewer than $5\%$ of all DSFGs sit at $z>4$ \citep{Dudzeviciute_2020} however, understanding precise volume densities at these epochs have been historically difficult due to issues leading from the negative k-correction of sub-mm detections and the lack of completeness in samples leading to $\sim1$ dex uncertainty on the volume density of DSFGs at $z\gtrsim4$ \citep{Long_2026}.
This means that our understanding of galaxy formation during the first $\sim$Gyr of cosmic time is limited during the epoch when massive galaxy growth is exponential. 

The limited understanding on the population and proliferation of DSFGs and their characterization in the early universe has highlighted a need for a multiwavelength study of the cSFRD from Cosmic Noon out into the Epoch of Reionization ($z>6$; \citealt{Barkana_2001,Robertson_2010}).
A complete, unbiased analysis would involve the direct investigation of stellar emission from newborn stars using the rest-frame ultraviolet (UV) and optical wavelengths, as well as the indirect measurement of stellar formation resulting from dust-enshrouded star-forming regions using the rest-frame far-infrared (FIR) and sub-millimeter regimes (see \citealt{Madau_2014} for review).
This analysis is required to accurately determine where DSFGs 
% within the first $\sim2$ Gyr of the cosmos 
fall along the star-forming main sequence \citep{Noeske_2007,Daddi_2007,Rodighiero_2011} as well as measure their physical characteristics (gas, dust, and stellar masses, star formation history (SFH), active galactic nuclei (AGN) content, etc.), information currently lacking due to the difficulties in characterizing (and thus confirming) an unbiased sample of DSFGs at $z>3$ \citep{Zavala_2018,Zavala_2021}.

The most dust-obscured and massive DSFGs were elusive in pre-James Web Space Telescope (JWST) optical and near-IR images due to their significant dust attenuation at short rest-frame wavelengths \citep{Casey_2014,Narayanan_2014,Wang_2019,Manning_2022}. 
This problem has compounds towards higher redshifts, for example, a substantial portion of all known $z > 6$ DSFGs were gravitationally lensed or serendipitously discovered in ALMA pointings of other high-$z$ targets, where such sources were deemed companions to the primary target, resulting in large uncertainties for any volume-based measurements \citep{Marrone_2018, Zavala_2018,Sun_2025, Fujimoto_2024,Inami_2022}.
% This problem has compounds towards higher redshifts, for example, nearly all known $z > 6$ DSFGs were gravitationally lensed or serendipitously discovered in ALMA pointings of other high-$z$ targets, where such sources were deemed companions to the primary target,
JWST/NIRCam has identified rest-frame optical emission of some $z\sim4$ DSFGs, but the most massive and dust-obscured galaxies at $z\ge5$ drop out of the shorter wavelengths filters in NIRCam ($\sim20\%$ at F150W; \citealt{Manning2025}) which introduces high uncertainties in photometric redshift estimates and other physical properties, hinting at an underestimation in the volume density of DSFGs at $z\ge5$ \citep{Akins_2023,McKinney_2025}.

% It is furthermore a concern that $z \sim 3-6$ DSFGs may contaminate $z > 7 - 10$ JWST objects. 
% This contamination is partly caused by both populations having similar, and uncertain, number densities \cite{McKinney_2023b}.

The leading observatory that can characterize DSFGs at $z>3$ is the Atacama Large (Sub)Millimeter Array (ALMA).
ALMA provides the sensitivity and angular resolution to efficiently observe the dust continuum of high-redshift galaxies at sub-mm wavelengths.
% due to its sensitivity and spatial resolution.
Several ALMA extragalactic deep fields have been carried out so far at increasing depths (e.g., GOODS-S; \citealt{Franco_2018} HUDF; \citealt{Dunlop_2017} SDXF; \citealt{Hatsukade_2016} ASPECS in HUDF/XDF; \citealt{Aravena_2016,Walter_2020})
% MORA; \citep{Casey_2021,Zavala_2021}) , and ExMORA; \citealt{Long_2026}) 
yet cover relatively small areas, thus enabling us to catch only a small glimpse of the expected population of DSFGs at $z > 3$. 
Larger surveys such as MORA \citep{Casey_2021,Zavala_2021} and ExMORA \citep{Long_2026} have improved on the population of IR-luminous galaxies at $z > 3$, yet find only a handful of the heavily dust-obscured objects that are known to exist at high redshift \citep[e.g.,][]{Bing_2026}.
Targeted ALMA surveys such as ALPINE \citep{Lefevre_2020,Bethermin_2020,Faisst_2020} and REBELS \citep{Bouwens_2022} have used ALMA to constrain the gas and dust properties of UV-selected galaxies at $z\sim4-8$, (i.e., sources on the high mass end of galaxies at their respective epochs) and have shown that these sources may also be significantly dust enriched \citep{Fudamoto_2020,Pozzi_2021,Gruppioni_2020,Inami_2022,Faisst_2022,Mauerhofer_2023,Burgarella_2025,Algera_2026}. 
Curiously, several galaxies were identified 
serendipitous to UV/optical imaging located near the primary target in these surveys \citep{Franco_2018,Gruppioni_2020,Romano_2020,Loiacono_2021,Fudamoto_2021,Van_Leeuwen_2024}.
These sources have bright [CII] emission and/or far-IR continuum but are undetected in rest-frame optical imaging.
These ``optically dark'' galaxies suggest that there is a hidden population of DSFGs at $z>4$ that may play a key role in early structure growth \citep{ Zou_2026}
% that are largely unknown 
and could contribute up to 10-25$\%$ to the $z>6$ cSFRD \citep{Fudamoto_2021}.

We therefore set out to perform a large, untargeted, and therefore unbiased, ALMA blank field survey to constrain the key populations of DSFGs and ``optically dark'' galaxies at $z>3$ with the \textit{COSMOS High-$z$ ALMA-MIRI Population Survey} (CHAMPS) program. 
Outlined by \citealt{Faisst_2026}, CHAMPS is an ALMA Large Program (\#2023.1.00180.L  PI: A. Faisst) designed to take sub-mm observations of the overlapping JWST MIRI+NIRCam footprint in the COSMOS field with the broad goal to understand the role dust played on the formation and growth of galaxies in the early universe by estimating the history of dust obscured star formation out to $z \sim 9$ (as well as other survey goals as described by \citealt{Faisst_2026}; see also \citealt{Zavala_2026}).

In this paper we present the data reduction, source detection methodology, source catalog, and first results of the CHAMPS program. 
The paper is organized as follows: In Section \ref{sec:observations}, we present the observational setup of CHAMPS along with the ancillary data in the COSMOS field that was used in our analysis. 
Section \ref{sec:reduction} covers the reduction of the ALMA CHAMPS data
% , including the calibration, the \texttt{CASA tclean} function along with its parameters, 
and several statistical measurements on the final reduced mosaic. In Section \ref{sec:source-detection}, we introduce the methods used to identify sources in the resulting mosaic.
In Section \ref{sec:catalog}, we present the resulting catalogs with total flux measurements, source number density, completeness,
% and acquire total flux measurements, estimate the number density, completeness, 
and provide a discussion of the limitations found in each catalog.
Section \ref{sec:properties} provides an analysis of the properties of the sample, covering the redshift distribution, stellar masses, and possible AGN candidates identified in CHAMPS and we conclude in Section \ref{sec:summary}.
When relevant, we use a standard $\Lambda$CDM cosmology with $H_0 = 70\,{\rm km\,s^{-1}\,Mpc^{-1}}$, $\Omega_\Lambda = 0.7$, and $\Omega_{\rm m} = 0.3$ and a Chabrier initial mass function \citep[IMF;][]{Chabrier_2003}. Magnitudes are given in the AB system \citep{Oke_1974}. 

\section{Observations}\label{sec:observations}
\begin{figure*}[t]
    \begin{center}
        \includegraphics[trim=0mm 0mm 0mm 0mm, clip, width=2\columnwidth]{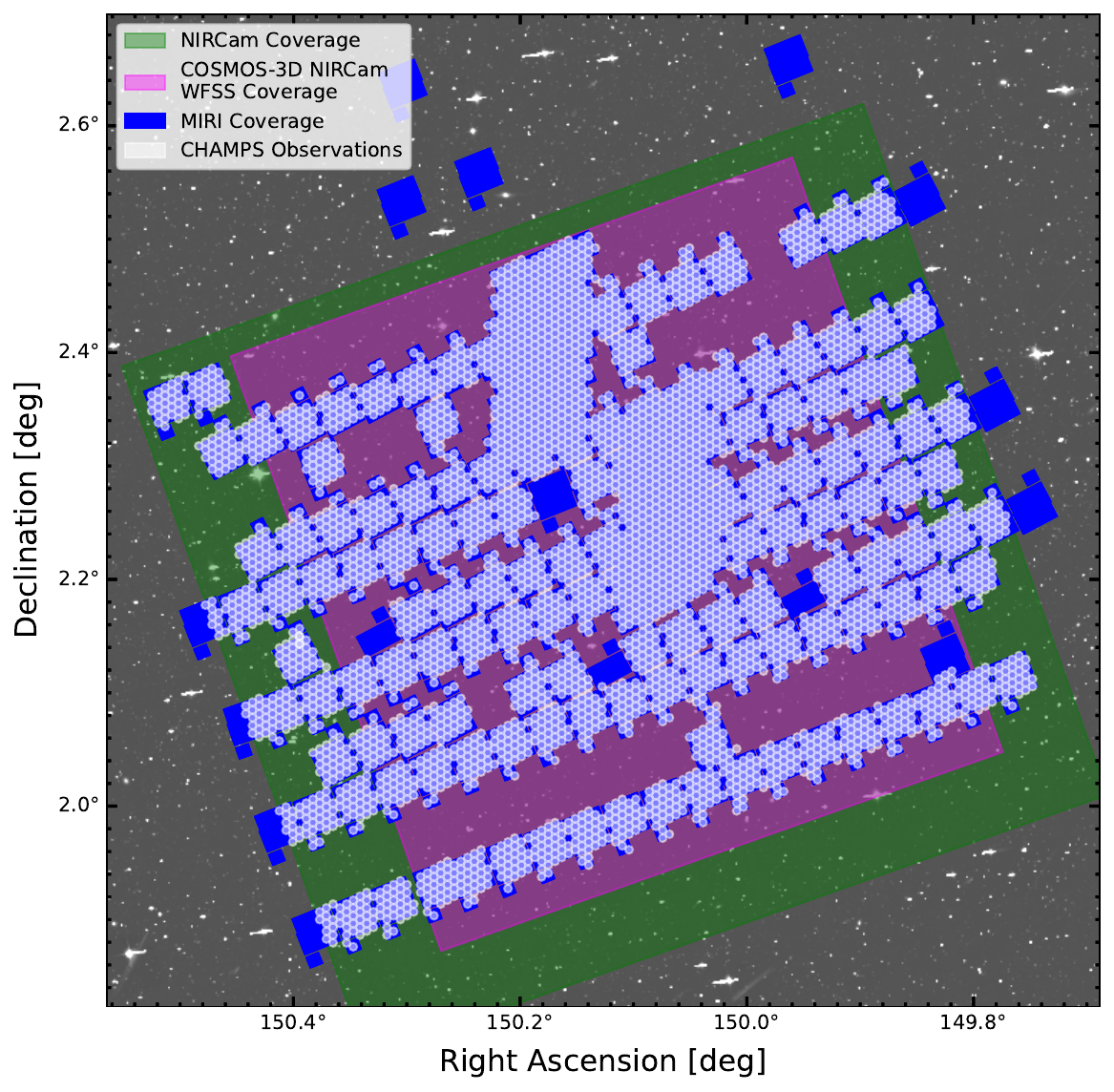}
         \caption{The layout of ALMA observations taken by CHAMPS. Background: HST ACS I-band (F814W) mosaic observations of COSMOS \citep{Koekemoer_2007}. CHAMPS footprint outlined in white with the MIRI coverage of COSMOS-Web and PRIMER outlined in blue. NIRCam coverage of COSMOS-Web is shown in green and NIRCam WFSS coverage from COSMOS-3D is displayed in pink. CHAMPS is designed to maximize overlap between NIRCam and MIRI and the remaining pointings that CHAMPS doesn't cover were those that were changed due to guide star failures.}\label{fig:observations}
    \end{center}
\end{figure*}

% CHAMPS, the {\it COSMOS High-$z$ ALMA-MIRI Population Survey} (\#2023.1.00180.L  PI: A. Faisst), is a 145-hour ALMA Cycle 10 band 6 Large Program designed to observe the overlapping 0.2 deg$^2$ of JWST MIRI + NIRCam footprints of COSMOS-Web, PRIMER, and COSMOS-3D in the COSMOS field at 1.2 mm ($\nu = 250$ GHz) (Figure \ref{fig:observations}). 
CHAMPS, the {\it COSMOS High-$z$ ALMA-MIRI Population Survey}, is a 145-hour ALMA Cycle 10 Large Program in band 6 at 1.2 mm ($\nu = 250$ GHz) designed to observe the overlapping 0.2 deg$^2$ of JWST/NIRCam$+$MIRI observations ($1-8\,{\rm \mu m}$) of the largest area JWST surveys in the COSMOS field, including COSMOS-Web (PID $\#$1727, PIs: J. Kartaltepe \& C. Casey, \citealt{Casey_Kartaltepe_2023}), PRIMER/COSMOS (PID $\#$1837, PI: J. Dunlop), and benefits from overlapping coverage in COSMOS-3D (PID $\#$5893, PI: K. Kakiichi, \citealt{Koki_2024}) (see Figure \ref{fig:observations}). 
As the largest ALMA blank-field survey to date,
CHAMPS has a total of 4800 pointings split between 32 different Scheduling Blocks (SBs) where each SB consists of 150 pointings. 
The total on-source integration time for each SB ranges from 54.4 seconds per pointing to 90.7 seconds per pointing,
the range of which resulting from the change of conditions in precipitable water vapor (PWV) at the time of observation, where increased PWV resulted in an increased integration time to achieve our desired root-mean-square (RMS).
Observations were taken with an average PWV of $1.3 \text{ mm}$, and a range of $0.29 \text{ mm} < \text{PWV} < 2.6 \text{ mm}$ (see Table \ref{table:SB-observations} for more details). 

Our observations are optimized to have a relatively constant RMS of $\sim0.1$ mJy across the full mosaic, equating to placement of the pointing centers with a $2 \times$ Nyquist sampling ($\sim24.66\arcsec$) spacing. 
This resulted in a uniform RMS noise of 146 $\mu$Jy beam$^{-1}$ and the majority of the CHAMPS survey (90$\%$) with an RMS at or below 200 $\mu$Jy beam$^{-1}$ for depths of $5\sigma$ (Figure \ref{fig:RMS}), discussions of which are continued in Section \ref{sec:reduction}.

\begin{table*}[t]
\centering
\begin{tabular}{c|cccccccc}

\midrule\midrule
\textbf{SB}	&	\textbf{Date(s) [2024]} &	\textbf{Config.}	&	\textbf{Time Per Pointing [s]} &	\textbf{PWV	[mm]} &	\textbf{Synth. Beam	Size [$\arcsec$]} & \textbf{\boldmath$\langle$RMS\boldmath$\rangle$ [$\mu$Jy]} \\
% \hline
\Xhline{1pt}
S1	&	Oct 30	-	Nov 1	&	C-2	&	54.43	&	1.52	&	1.08	& $146^{+4}_{-7}$ \\
S2	&	Sep 9	&	C-3	&	54.43	&	0.85	&	0.63	& $133^{+4}_{-9}$ \\
S3	&	Mar 8	-	 21	&	C-1	&	54.43	&	1.19	&	1.10	& $148^{+4}_{-10}$ \\
S4	&	Jan 24	-	 25	&	C-3	&	72.58	&	2.04	&	0.66	& $155^{+3}_{-10}$ \\
S5	&	Apr 5	-	 17	&	C-2	&	90.72	&	2.08	&	0.93	& $127^{+2}_{-7}$ \\
S6	&	Mar 7	-	 8	&	C-1	&	54.43	&	1.04	&	1.11	& $151^{+4}_{-7}$ \\
S7	&	Sep 21	-	 27	&	C-3	&	54.43	&	0.5	&	0.67	& $126^{+8}_{-6}$ \\
S8	&	Sep 5	-	 7	&	C-4	&	54.43, 72.57	&	0.45, 0.49	&	0.60	& $107^{+6}_{-9}$ \\
S9	&	Jan 8	-	 23	&	C-4, C-3	&	54.43	&	1.17	&	0.50	& $139^{+4}_{-6}$ \\
S10	&	Jan 29	-	Mar 6	&	C-3, C-1	&	54.43	&	1.89	&	0.66	& $163^{+3}_{-7}$ \\
S11	&	Sep 20	- 21	&	C-3	&	54.43	&	0.73	&	0.72	& $128^{+4}_{-7}$ \\
S12	&	Sep 13	-	Oct 30	&	C-3, C-2	&	54.43, 72.57	&	0.38, 0.98	&	0.73	& $124^{+4}_{-18}$ \\
S13	&	Sep 21	-	 27	&	C-3	&	54.43	&	0.29	&	0.65	& $106^{+5}_{-6}$ \\
S14	&	Oct 18	-	 29	&	C-3, C-2	&	54.43	&	0.75	&	0.73	& $133^{+5}_{-5}$ \\
S15	&	Nov 19	-	 28	&	C-1	&	54.43	&	2.08	&	1.09	& $138^{+2}_{-6}$ \\
S16	&	Sep 15		&	C-3	&	54.43	&	0.33	&	0.62	& $113^{+7}_{-6}$ \\
S17	&	Apr 24	-	 25	&	C-3	&	54.43	&	1.52	&	0.67	& $145^{+2}_{-5}$ \\
S18	&	Mar 8	-	 18	&	C-1	&	54.43	&	1.86	&	1.16	& $155^{+4}_{-5}$ \\
S19	&	Nov 15	-	 19	&	C-1	&	72.58	&	2.31	&	1.09	& $144^{+3}_{-7}$ \\
S20	&	Oct 16	-	 17	&	C-3	&	72.58	&	1.93	&	0.63	& $132^{+2}_{-5}$ \\
S21	&	Sep 11	-	 15	&	C-3	&	54.43	&	1.54	&	0.63	& $135^{+5}_{-7}$ \\
S22	&	Apr 28	-	May 5	&	C-3	&	90.72	&	2.44	&	0.65	&$ 145^{+8}_{-10}$ \\
S23	&	Sep 18	-	 20	&	C-3	&	54.43	&	0.45	&	0.70	& $123^{+3}_{-6}$ \\
S24	&	Oct 17		&	C-3	&	54.43	&	0.82	&	0.67	& $142^{+4}_{-10}$ \\
S25	&	Oct 4	-	 15	&	C-3	&	54.43	&	0.97	&	0.65	& $134^{+5}_{-7}$ \\
S26	&	Mar 22	-	Apr 2	&	C-1, C-2	&	54.43	&	1.82	&	1.08	& $150^{+9}_{-11}$ \\
S27	&	Nov 12	-	 15	&	C-1	&	54.43	&	1.32	&	1.06	& $149^{+2}_{-5}$ \\
S28	&	Sep 16		&	C-3	&	54.43	&	0.51	&	0.68	& $122^{+5}_{-5}$ \\
S29	&	May 9	-	Aug 29	&	C-3, C-4	&	54.43	&	1.3	&	0.47	& $145^{+4}_{-7}$ \\
S30	&	Sep 10		&	C-3	&	54.43	&	0.76	&	0.67	& $122^{+1}_{-3}$ \\
S31	&	Aug 29	-	Sep 6	&	C-4	&	54.43	&	0.44	&	0.46	& $139^{+6}_{-7}$ \\
S32	&	Apr 26	-	 28	&	C-3	&	72.58	&	2.6	&	0.65	& $163^{+5}_{-7}$ \\
\midrule\midrule
\end{tabular}
\caption{Table displaying what dates and configurations each Schedule Block (SB) was observed in. Also displayed are the average integration time per pointing, the precipitable water vapor (PWV) at the time of observation, the synthesis beam size, and the median root-mean-square (RMS) for each SB with $\pm1\sigma$ confidence values are included.}
\label{table:SB-observations}
\end{table*}

Observations are focused towards the high-frequency end of ALMA's band 6 with a local oscillator frequency tuning centered at 250 GHz ($\lambda$ = 1.2 mm). 
CHAMPS utilizes four $1875$ MHz-wide spectral windows centered at 241 GHz, 243 GHz, 257 GHz,  and 259 GHz, covering a total bandwidth of 7.5 GHz.
Band 6 observations are ideal for the selection of dust-obscured $z>3$ galaxies \citep[][]{Casey_2014,Hodge+da_Cunha_2020}. 
Due to the negative $k-$correction and the typical dust spectral energy distribution (SED) shape, galaxies at redshifts $z=3-6$ are brightest in this frequency window \citep[][]{Blain_2002}, hence increasing the source detectability of CHAMPS.

In order to optimize the available observing time in band 6, observations took place over multiple months during ALMA Cycles 10 and 11, from January to November 2024 in configurations C-1, C-2, C-3, and C-4.
Observations used 41-48 antennae with a maximum baseline of $\sim$784 meters resulting in a varying synthesis beam size ranging from $0.5\arcsec \lesssim \theta_{\rm{FWHM}} \lesssim 1.1\arcsec$ across our observations (see Table \ref{table:SB-observations} and Figure \ref{fig:beamsize}).

\subsection{Ancillary Data}\label{sec:ancillary}

CHAMPS was designed to be completely covered by JWST NIRCam+MIRI observations (1 - 8 $\mu$m) through several large JWST surveys.
COSMOS-Web 
% (PID $\#$1727, PIs: J. Kartaltepe \& C. Casey; \citealt{Casey_2023}) 
is a JWST Cycle 1, 255 hour, survey with NIRCam imaging in four filters (F115W, F150W, F277W, and F444W) and MIRI imaging in one (F770W).
Both NIRCam and MIRI observations were taken in parallel, resulting in the striped coverage of MIRI (and thus the resulting coverage of CHAMPS) across the COSMOS field (Figure \ref{fig:observations}). 
Note that a handful of pointings from COSMOS-Web had guide star failures in their initial configuration and were re-observed with a different position angle. 
CHAMPS pointings were allocated to regions with MIRI coverage at full depth, hence the few MIRI pointings shown in Figure \ref{fig:observations} without CHAMPS coverage.

PRIMER/COSMOS 
% (PID $\#$1837, PI: J. Dunlop; \citealt{Dunlop_2021}) 
is a JWST Cycle 1, 187 hour survey with deep coverage on portions of both the COSMOS and UDS fields with NIRCam in 8 filters (F090W, F115W, F150W, F200W, F277W, F356W, F444W, and F410M) as well as MIRI in two filters (F770W and F1800W).
PRIMER observations in COSMOS focus on the HST CANDELS-COSMOS footprint \citep{Grogin_2011,Koekemoer_2011} and make up the two contiguous rectangles of MIRI coverage seen in Figure \ref{fig:observations}.

COSMOS-3D 
% (PID $\#$5893, PI: K. Kakiichi; \citealt{Koki_2024}) 
is a JWST Cycle 3 survey with NIRCam wide field slitless spectroscopic (WFSS) observations in F444W and direct NIRCam imaging in three filters (F115W, F200W, and F356W), as well as MIRI parallel imaging in two filters (F1000W and F2100W) covering the inner 60\% of the COSMOS-Web area. 
The COSMOS-3D MIRI parallels were designed to overlap with the COSMOS-Web MIRI parallels, resulting in three filter coverage in those areas.
% As COSMOS-3D was approved after CHAMPS was benefits from having decent overlapping coverage with COSMOS-3D ($\sim XX\%$), 
Unfortunately, there are portions of COSMOS-3D MIRI observations that do not have CHAMPS coverage, this is due to CHAMPS being awarded time before COSMOS-3D was accepted.
We used the COSMOS-3D [O III] catalog \citep{Meyer_2026} for cross matching redshift estimates in our analysis of CHAMPS sources (see Section \ref{sec:redshift} for details).

In addition to JWST observations, there is a wealth of ground- and space-based optical/near-IR data in the COSMOS field that has been assembled into a comprehensive multiwavelength catalog (COSMOS2025; \citealt{Shuntov_2025}).
Including the aforementioned JWST/NIRCam+MIRI imaging from COSMOS-Web and PRIMER, COSMOS2025 also includes Hubble/F814W imaging \citep{Koekemoer_2007,Koekemoer_2011}, the Spitzer
Cosmic Dawn Survey \citep{Euclid_2022,Euclid_2025},
Subaru Telescope Hyper Suprime-Cam (HSC) imaging \citep{Aihara_2022}, and UltraVISTA imaging \citep{McCracken_2012}. 
COSMOS2025 combines these ground and space-based data to derive photometric measurements of NIRCam-detected sources using both fixed-aperture photometry and profile fitting for 37 bands.
With photometry, photometric redshifts, and morphological and physical parameters for nearly 800,000 galaxies that have been identified in the COSMOS-Web imaging, COSMOS2025 was used throughout the CHAMPS catalog creation through source detection and to obtain galaxy properties.

We procure spectroscopic redshifts from various sources, primarily from the COSMOS Spectroscopic Redshift Compilation \citep{Khostovan_2026} (including \citealt{Lilly_2007, Trump_2007, Coil_2011, Casey_2012, Krogager_2014, Scoville_2015, Casey_2015,Kartaltepe_2015,Kriek_2015,Onodera_2015, Onodera_2016, Nanayakkara_2016, Casey_2017, Hasinger_2018, Jin_2019, Kashino_2019,Masters_2019,Wisnioski_2019, Shah_2020, Polletta_2021,vanDerWel_2021, Chen_2022,Jin_2022,Lilly_2023, Cooper_2023, Cooper_2024, Epinat_2024, Gentile_2024,Sillassen_2024, Forrest_2025, DESI_Collaboration_2026, Lertprasertpong_in_prep, Vanderhoof_in_prep}). 
This dataset compiles 20 years of spectroscopic observations in the COSMOS field, listing redshift estimates from over 130 different programs, catalogs, survey webpages, publications, and individual PI's.
Further spectroscopic redshift values were pulled from the DAWN JWST Archive v4.4 (DJA) \citep{de_Graaff_2024,Heintz_2025,Valentino_2025,Gillman_2026}.
DJA is an initiative of the Cosmic Dawn Center (DAWN) 
that presents a public repository of JWST galaxy data and spectroscopic redshift estimates, reduced with \texttt{grizli} \citep{brammer_2023_grizli} and \texttt{msaexp} \citep{Brammer_2023_msaexp}.
Finally, we obtained spectroscopic observations from various programs across COSMOS that are missing from the aforementioned compilations. 
The following spectroscopic programs have NIRSpec/MSA observations in the COSMOS field that were used in the cross matching of CHAMPS sources: CAPERS (GO$\#$6368, PI: M. Dickinson), Mirage or Miracle?$\ $(MoM; GO$\#$5224, PI: P. Oesch), EMBER (GO$\#$7076, PI: H. Akins), and ZENITH (GO$\#$7417,  PI: C. Casey).

The COSMOS field is also well probed across X-ray, millimeter, and radio wavelengths. 
X-ray and radio data are primarily used to discern the presence of powerful AGN, and to constrain whether AGN emission contributes to the 1.2 mm flux and/or the galaxy’s multiwavelength SED, while sub-mm data were used to validate sources identified in the mosaic and refine the dust SED models while providing accompanying photometric redshifts. 
X-ray data was obtained through Chandra observations from COSMOS-Legacy survey \citep{Civano_2012, Civano_2016}. 
% X-ray and radio data are primarily used to discern the presence of powerful AGN, and to quantify whether AGN emission contributes to the 1.2 mm flux and/or the galaxy’s multiwavelength SED. 
% For these analyses, we use X-ray data from Chandra through the COSMOS-Legacy survey \citep{Civano_2012, Civano_2016}. 
Millimeter and sub-millimeter datasets include SCUBA-2 maps at 450 $\mu$m + 850 $\mu$m (STUDIES, \citealt{Wang_2017}; S2COSMOS, \citealt{Simpson_2019}) as well as the S2COSMOS based dataset SCUBADive \citep{McKinney_2025}.
More sub-millimeter coverage include ALMA observations at 2 mm (Ex-MORA, \citealt{Casey_2021,Zavala_2021,Long_2026}), and archival ALMA observations identified in the A3COSMOS survey \citep{Liu_2019,Adscheid_2024}. 
Radio data in our analysis include deep continuum coverage by the VLA at 3 GHz and 10 GHz from the COSMOS-XS survey \citep{van_der_Vlugt_2021}, the full-field 3 GHz VLA-COSMOS Large Project \citep{Smolcic_2017}, and the MeerKAT MIGHTEE-COSMOS 1.28 GHz survey \citep{Heywood_2022}.

\section{Data Reduction}\label{sec:reduction}

\subsection{Calibration}
All 32 SBs making up the CHAMPS dataset were calibrated using the CASA-ALMA version 6.6.1.17 pipeline \citep{Hunter_2023}.
% using scripts provided from the ALMA array.
The automatically-generated calibration reports (Weblogs) were examined to check for suspicious behaviors (e.g., unexpected low gain or high system temperatures, ensuring resolution and beam sizes fit within desired ranges, and locating any faulty antennae). 
% With no major issues identified, we continued onto the reduction.
% and identified no problems with our data.
Each SB was split into several Measurement Sets (MSs) due to the number of pointings and size of each dataset with a total number of 106 MSs in the CHAMPS dataset.
These MSs are the non-continuum subtracted datasets for each SB and are the files used for \texttt{tclean}. 
No major issues were identified during this visual inspection process.
% With no major issues identified, we continued onto the reduction.

\begin{figure*}[t]
    \begin{center}
        \includegraphics[trim=0mm 0mm 0mm 0mm, clip, width=2\columnwidth]{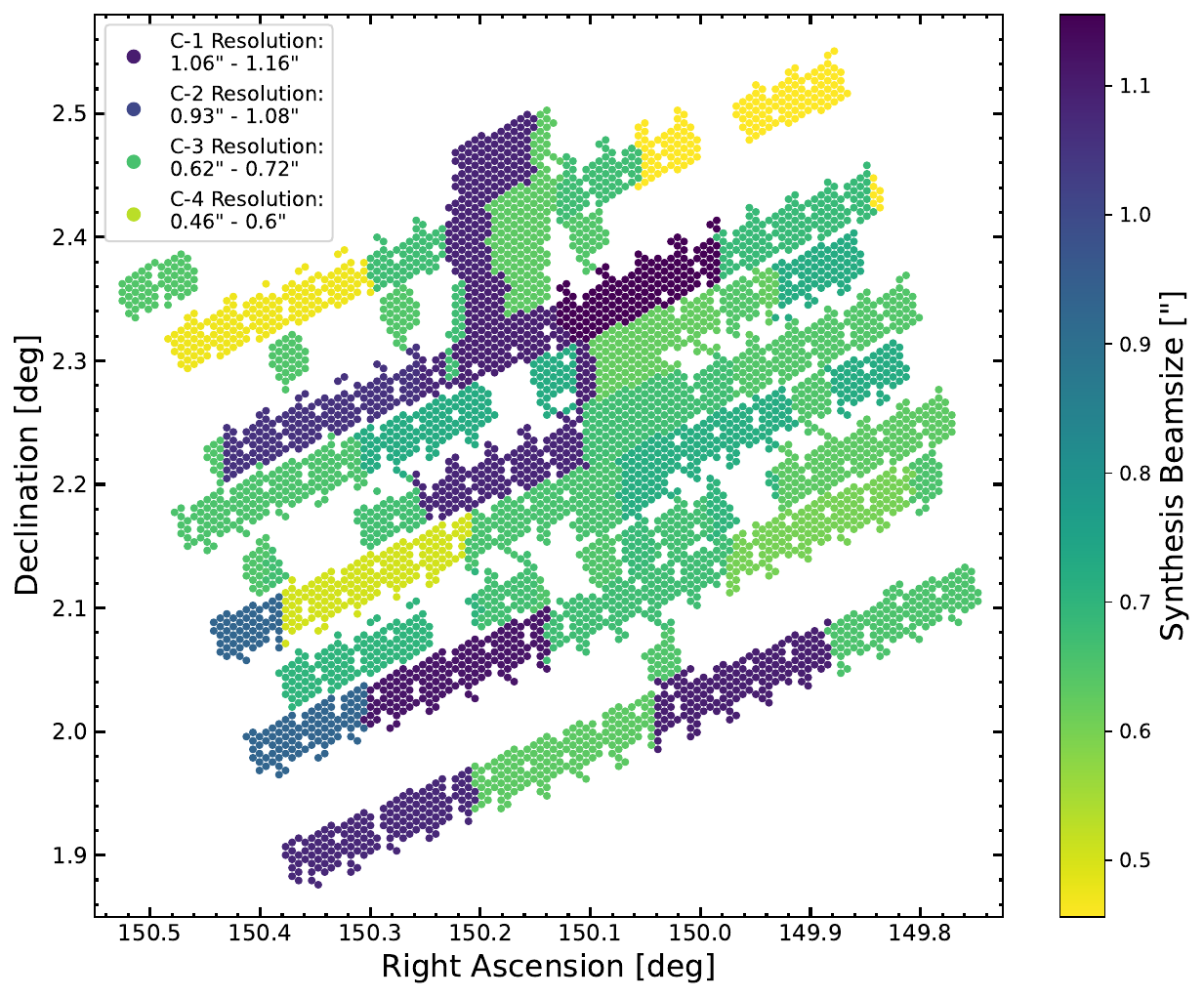}
         \caption{Map showing the synthesis beam size of CHAMPS pointings. The beam size for the pointings in CHAMPS ranges from $0.46\arcsec - 1.16\arcsec$ due to observations being taken across multiple configurations (see more details in Table \ref{table:SB-observations}). The corresponding beam size ranges for each configuration can be seen in the legend tracking the values displayed in the figure. We chose a restoring beam size of $\theta_\text{res} = 1.1\arcsec \times 1.1\arcsec$ for the reduction as it follows the largest beam size observed by the CHAMPS observations.
         % Varying synthesis beamsize in the CHAMPS mosaic. 
         }\label{fig:beamsize}
    \end{center}
\end{figure*}

\subsection{Imaging}
CHAMPS observations were imaged using the CASA-ALMA version 6.6.1.17 \texttt{tclean} command, 
a task based on the \texttt{CLEAN} algorithm \citep{Hogbom_1974}, which is utilized to reconstruct a model image from interferometric data.
To achieve a continuous mosaic across the CHAMPS footprint, we imaged the entire mosaic at once (all 4800 pointings making up the 32 SBs).
% Without altering the data, this would lead to heavy computing power and long wait times with \texttt{tclean}.

To optimize imaging, each MS was split in frequency and time bins using CASA-ALMA's \texttt{split} command.
Splitting large observations can lower the total amount of data going through imaging and greatly reduces the processing power and computation time without losing vital information.
As the goal of this work is to identify sources in the continuum (the reduction of CHAMPS spectral cubes will be discussed and presented in a forthcoming paper) we can perform this splitting without affecting our science goals.
Time averaging was conducted with time bins of 10 seconds, a value small enough to not impact the data quality but large enough to lessen the processing power required for imaging.
We used 8-channel bins of 125 MHz for frequency averaging, as 8 bins is the maximum number of channels ALMA band 6 observations can receive before experiencing smearing \citep{Kepley_2023}.
Binning in frequency space has the added benefit of suppressing the noise while boosting the flux of the mosaic, increasing the chances of detecting faint sources.
This process was performed for each MS making up the dataset and returned an equal number of split MSs.
Splitting is the first step performed in imaging and ensures that double-binning does not occur for any parameter later down in our reduction pipeline.

After splitting, we concatenated the resulting split MSs into a \texttt{combined$\_$split$\_$MS} that includes the entire CHAMPS dataset into one MS via CASA-ALMA's \texttt{concat} command.
This was done for both convenience and to obtain the optimum parameters for a full field reduction.
An optimum physical cell size was identified to be $0.15\arcsec$ per pixel using the  \texttt{pickCellSize} command within the \texttt{analysisUtils} package \citep{Todd_2026} in CASA-ALMA on the \texttt{combined$\_$split$\_$MS}.

Observations were taken with multiple configurations, resulting in varying synthesis beam sizes throughout the field (see more in Section \ref{sec:observations}, Table \ref{table:SB-observations}, and Figure \ref{fig:beamsize}). 
To accommodate for this in the reduction, a uniform restoring beam size across the mosaic was set to be $\theta_\text{res} = 1.1\arcsec \times 1.1\arcsec$, which is equivalent to the largest synthesis beam size achieved by the CHAMPS observations.

When combining interferometric observations, \texttt{tclean} gives a weight to each baseline pair of antennas.
Measured in the \textit{u,v} plane, the baseline of an antenna pair is defined as the distance between the antennas as viewed by the source on the sky.
Weighting these distances allows for improvements in the dynamic range (the ratio between the strongest and weakest signals) and the ability to adjust the synthesized beam associated with the final mosaic.
The \texttt{weighting} parameter is typically adjusted in \texttt{tclean} to produce a higher sensitivity to extended sources or to account for noise variation between visibility samples.
As CHAMPS is a large blank-field survey that is meant to detect sources at low resolution, a \texttt{natural} weighting scheme was chosen.
Natural weighting gives equal weight to all baselines, resulting in the lowest noise level and largest resolution, with a downside of relatively high sidelobe levels.
As there are no unusually bright sources in the surveyed area, we do not need to be concerned with the sidelobes in imaging. 
Thus making \texttt{natural} weighting, corresponding to a robust value = 2, the ideal choice to optimize source signal-to-noise ratios (SNR).

% Finally, we centralized the reduction the \texttt{tclean} parameter, \texttt{phasecenter}, to be located on the most centralized pointing in our reduction (That of S12-54 located at ).

We then ran CASA-ALMA’s \texttt{tclean} on the \texttt{combined$\_$split$\_$MS} with the aforementioned parameters included. 
No cleaning, masking, or deconvolution was performed during the reduction, \texttt{tclean} was used purely as a mosaicing tool.
Deconvolution helps to remove the sidelobe effect of bright sources, as CHAMPS has no bright sources in its surveyed area, deconvolution would not change the resulting mosaic but would greatly increase the processing power required for the reduction.
As such, the resulting output for the reduced mosaic could be thought of as a ``dirty" map which, for a blank field survey, is acceptable for source detection.

\begin{figure}[t]
    \begin{center}
         \includegraphics[trim=0mm 0mm 0mm 0mm, clip, width=1\columnwidth]{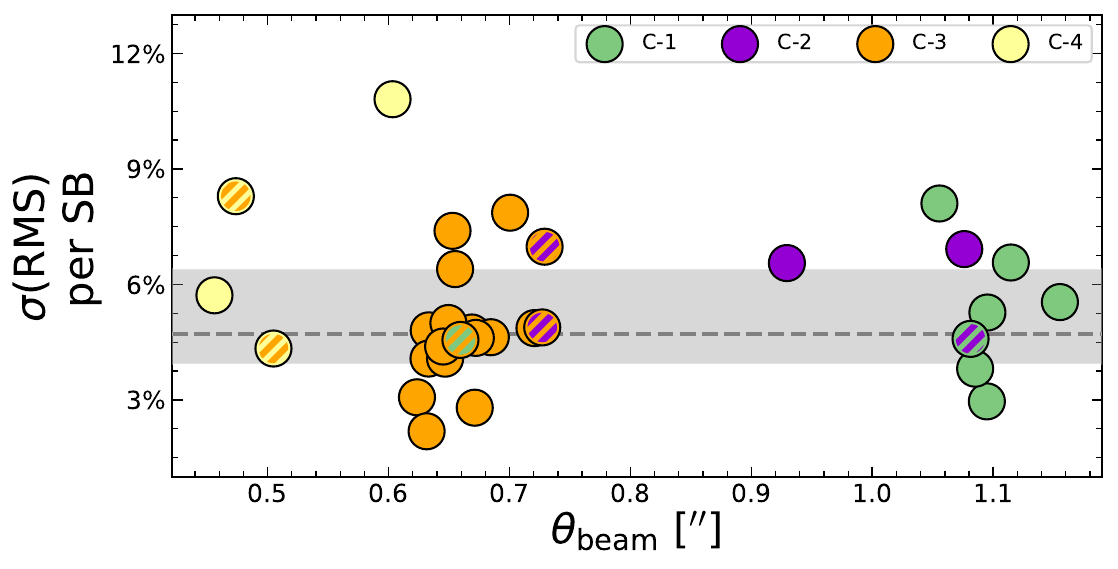}
         \includegraphics[trim=0mm 0mm 0mm 0mm, clip, width=1\columnwidth]{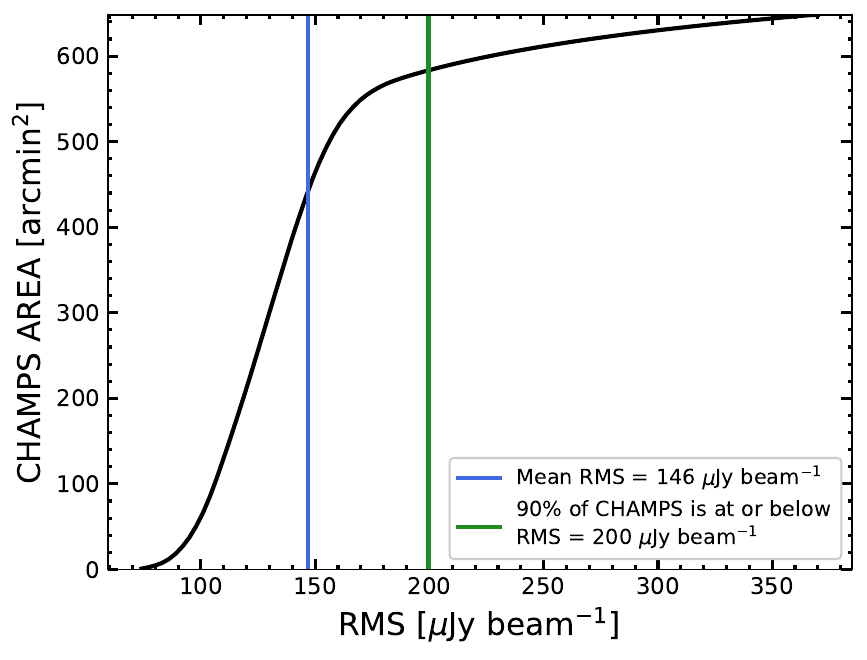}
         \caption{On the top, we show the variations in the RMS per SB as a function of synthesis beam size (per SB). 
         The gray band shows the $\pm1\sigma$ spread. 
         Symbols are color-coded by their respective ALMA configurations, SBs observed in two different configurations hold a combination of hatched colors from both configurations. 
         We find an RMS variation of $3-6\%$ in for SBs of similar beam size.
         On the bottom, we show the RMS cumulative distribution for CHAMPS (black solid). The average (mean) RMS for CHAMPS is measured at $146$ $\mu$Jy beam$^{-1}$ (blue vertical) and the majority of the CHAMPS survey (90$\%$) holds an RMS at or below 200 $\mu$Jy beam$^{-1}$ (green vertical) for depths of $5\sigma$.}
         \label{fig:RMS}
         % \caption{On the top, we show the RMS cumulative distribution for CHAMPS (black solid). The average (mean) RMS for CHAMPS is measured at $146$ $\mu$Jy beam$^{-1}$ (blue vertical) and the majority of the CHAMPS survey (90$\%$) has depths at or below 200 $\mu$Jy beam$^{-1}$ (green vertical). On the bottom, we show the histogram of pixel values in the signal-to-noise maps for CHAMPS (black histogram). The histogram is well fit by a standard normal (Gaussian) distribution, centered at 0, with $\sigma$ = 1 (red line), confirming Gaussian properties of the map’s noise. 
         % Above an SNR $\sim 4$ (black dotted), we see a divergence from the standard normal due to the presence of positive source peaks (teal shaded) which decouple fully from noise peaks (gray shaded) at an SNR $\sim6$.}
         % \label{fig:SNR-RMS}
    \end{center}
\end{figure}

With the outputs from \texttt{tclean}, we perform a primary beam correction using the CASA-ALMA command \texttt{impbcor}.
A primary beam correction is used to correct for the pointing sensitivity across the resulting mosaic and is necessary when determining accurate flux values.
% We used the primary beam corrected data for statistical analysis and source detection.
We use the primary beam corrected map ($M_\text{PBcorr}$) along with the primary beam map  ($M_\text{PB}$) to calculate the RMS map
% root-mean square (RMS) maps 
($M_\text{RMS}$) and SNR
% the signal-to-noise ratio (SNR) 
map ($M_\text{SNR}$) using the following relationships:
\begin{equation}
\begin{split}
    M_{\text{RMS}} = \text{RMS}_{\text{uniform}} / M_{\text{PB}},
    \\
    M_{\text{SNR}} = M_{\text{PBcorr}} / M_{\text{RMS}}, 
    % \\
    % M_{\text{Negative}} = (-1)\times\text{PB}_{\text{Dat}},
\end{split}\label{eq:RMS/SNR}
\end{equation}
where $\text{RMS}_\text{uniform}$ is the standard deviation of our non-primary beam corrected mosaic measured in Jy/beam.
These maps are used for quality testing the data and source detection.

\begin{figure}[t]
    \begin{center}
    \includegraphics[trim=0mm 0mm 0mm 0mm, clip, width=1\columnwidth]{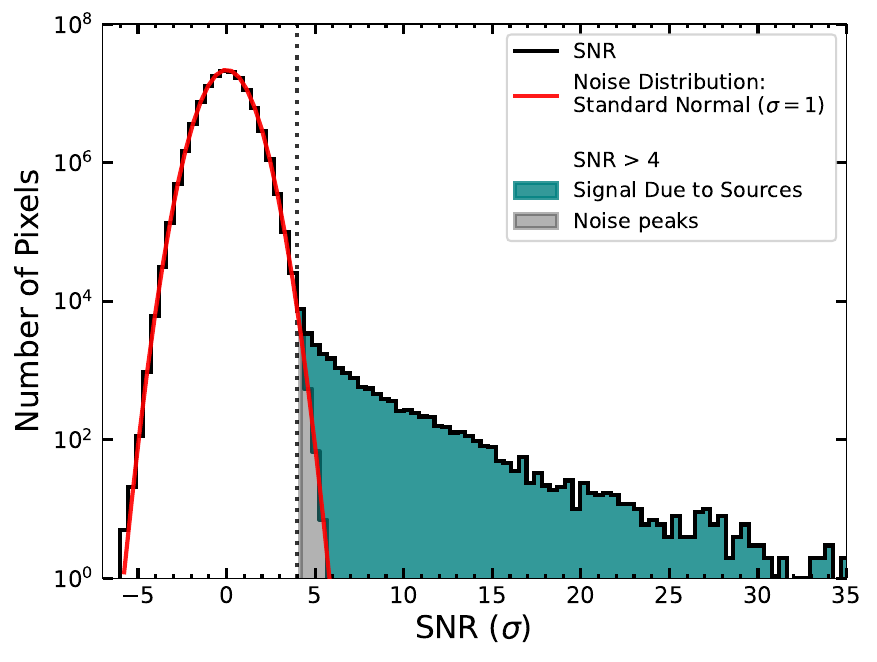}
    \caption{Histogram of pixel values in the signal-to-noise maps for CHAMPS (black histogram). The histogram is well fit by a standard normal (Gaussian) distribution, centered at 0, with $\sigma$ = 1 (red line), confirming Gaussian properties of the map’s noise. 
    Above an SNR $\sim 4$ (black dotted), we see a divergence from the standard normal due to the presence of positive source peaks (teal shaded) which decouple fully from noise peaks (gray shaded) at an SNR $\sim6$.}
    \label{fig:SNR}
    \end{center}
\end{figure}

% As expected, we observe that the RMS varies across the RMS map due to the configuration differences.
We observe the RMS varying across the RMS map due to differences in the PWV and configuration that each SB was taken in.
For SBs that were taken in the same configuration, we measure that the RMS varies from $\sim3 - 6\%$.
We find that the average RMS for the CHAMPS mosaic is measured at $146$ $\mu$Jy beam$^{-1}$ and that the majority of the CHAMPS survey (90$\%$) holds an RMS at or below 200 $\mu$Jy beam$^{-1}$ for depths of $5\sigma$ (Figure \ref{fig:RMS}).

\section{Source Detection}\label{sec:source-detection}

We used two methods, resulting in two catalogs, to identify sources in the CHAMPS mosaic.
The first method involves a blind search of the CHAMPS mosaic to identify sources corresponding to the point of their peak SNR.
This method is best-suited for identifying sources above an SNR $\ge 5\sigma$, but suffers from a high false-positive detection rate for sources when approaching lower SNRs.
The second method uses prior MIRI observations to identify sources in the CHAMPS mosaic.
Prior observations allows a deeper probe into the mosaic to identify faint sources that could otherwise have been mistaken for noise with a blind search, but could potentially bias the sample.
Both methods are discussed thoroughly in the following section.

\subsection{Blind Peak SNR Detections}\label{sec:blind_detections}
We identify source positions and flux densities corresponding to the point of their peak SNR for sources at or above 5$\sigma$. As shown in Figure \ref{fig:SNR}, the SNRs of the map pixels are well represented by a standard normal (Gaussian) centered about zero with $\sigma = 1$, above which positive source peaks diverge
from the distribution, indicating that the SNR $\ge 5 \sigma$ threshold is a reasonable threshold for low source contamination.
We searched the mosaic to look for clumps of pixels that are above this threshold by using the peak location function \texttt{find$\_$peaks()} in \texttt{photutils} version 1.3.0, an Astropy package designed for detection of astronomical sources \citep{Bradley_2026}.
% Overall, we identified 385 sources in CHAMPS at or above $5\sigma$ significance.

\begin{figure*}[t]
    \begin{center}
        \includegraphics[trim=0mm 0mm 0mm 0mm, clip, width=2\columnwidth]{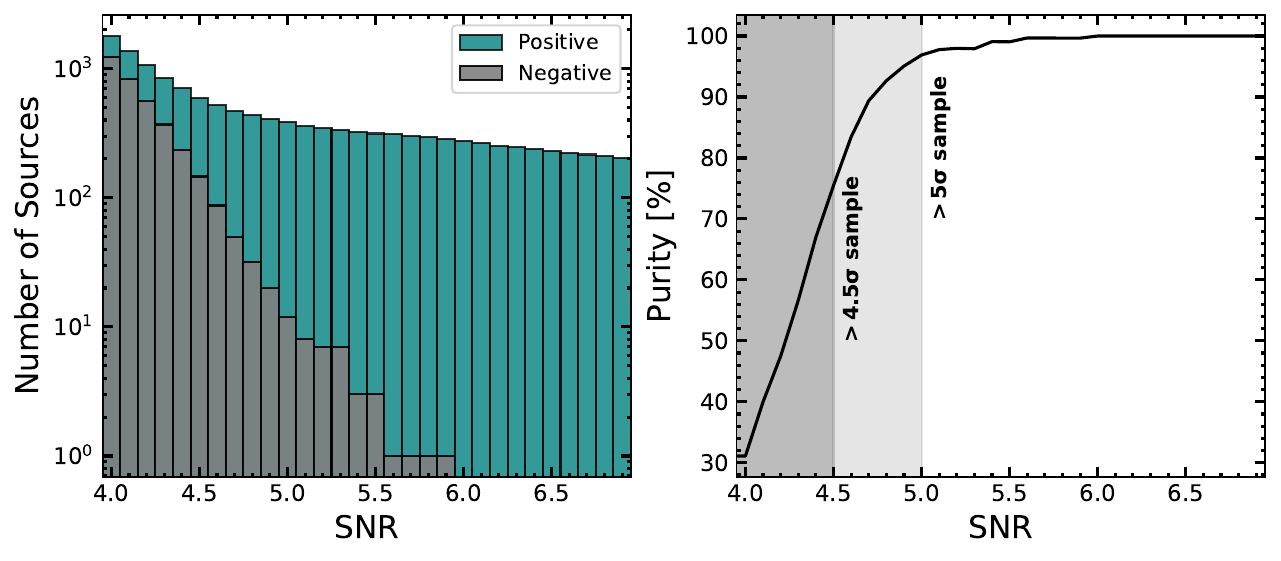}
         \caption{[left] Positive (teal histogram) and negative (grey histogram) sources identified with a blind search across all CHAMPS observations. [right] Purity of blind detections with increasing SNR represented by Equation \ref{eq:purity}. The $5\sigma$ sample used for our analysis is highlighted in the white region, while the $4.5\sigma$ sample is shown in the gray block. We reach a purity of $96.9\%$ above an SNR of $\ge5\sigma$ ($N_\text{pos} = 385$; $N_\text{neg} = 12$).}\label{fig:purity}
    \end{center}
\end{figure*}

To account for noise fluctuations in the reduction, the same process was performed on the negative of the SNR map (that is taking the SNR map and multiplying it by $-1$). 
As Gaussian noise can lead to negative detections, this process is helpful in identifying how many noise peaks we can expect to appear across the mosaic. 
We display the number of positive and negative detections as a function of SNR in the left panel of Figure \ref{fig:purity}.
These negative detections are, by definition, not real and represent how the map noise can vary across the mosaic and how it increases near the edge of an observation. 

The difference between positive and negative sources is a proxy for the expected number of real sources and can be thought of as a purity metric, defined as:
\begin{equation}
    P = \frac{N_\text{pos} - N_\text{neg}}{N_\text{pos}},
\end{equation}\label{eq:purity}

\noindent where $N_\text{pos}$ and $N_\text{neg}$ are the numbers of positive and negative detections respectively. 
The purity of a sample can help describe the confidence that a source identified at any given SNR is a real object and not a noise peak. %that was flagged mistakenly.
The purity of our detections as a function of SNR is displayed in the right panel of Figure \ref{fig:purity}, where we have identified the
purity for SNR $\ge5 \sigma$ sources ($N_\text{pos} = 385$; $N_\text{neg} = 12$) to be $P (\ge5\sigma) = 96.9\%$.

\begin{figure}[h!]
    \begin{center}
        \includegraphics[trim=0mm 0mm 0mm 0mm, clip, width=1\columnwidth]{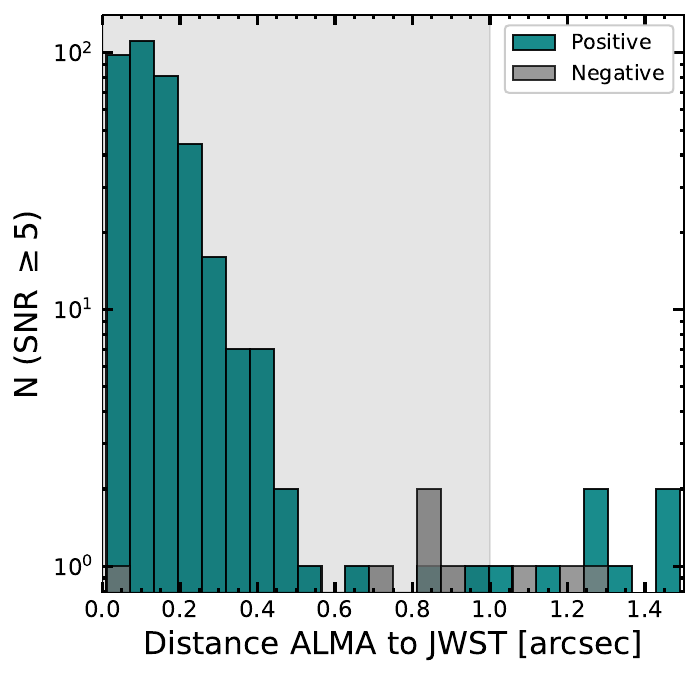}
         \caption{Distance from $>5\sigma$ ALMA detections to the nearest JWST/NIRCam counterpart identified \cite{Shuntov_2025}. This sample consists of both positive (teal histogram) and negative (black histogram) SNR $> 5\sigma$ detections. The gray region highlights the sources that return a match within a search radius of $\Delta r(\text{NIRCam}) = 1''$.} \label{fig:source_distances}
    \end{center}
\end{figure}

To identify how many of our sources were detected in previous catalogs, we cross-matched the positions for all SNR $\ge5\sigma$ sources with vetted detections identified in prior catalogs, most notably the NIRCam-based detections in COSMOS using the COSMOS2025 catalog \citep{Shuntov_2025}. 
Searching within $\Delta r(\text{NIRCam}) =  1\arcsec$, we found that most positive sources matched (370/385), while only a few negative sources returned a match (5/12) (Figure \ref{fig:source_distances}). 
If multiple JWST detections were identified within this range, favorability was given to the source with the reddest color ($m_{\text{F277W}} - m_{\text{F444W}}$).
% This returned 370/385 $> 5 \sigma$ positive peaks within the $\le 0.6''$ NIRCam search radius, while only returning 5/12 negative $>5 \sigma$ peak.
We define a ``modified purity'' metric for the positive and negative peaks that meet this criteria (SNR $\ge 5 \sigma$; $\Delta r(\text{NIRCam}) = 1''$), returning a results of $P(\ge5\sigma)_{\rm{modified}} = 98.65\%$.
These results provide a high confidence that CHAMPS SNR $\ge5\sigma$ source candidates that overlap with a detection in COSMOS2025 are likely real objects and are not noise fluctuations.

Further cross matching was also performed with the sub-mm and radio datasets, A3COSMOS \citep{Liu_2019, Adscheid_2024}, ExMORA \citep{Casey_2021, Zavala_2021, Long_2026}, SCUBADive \citep{McKinney_2025}, VLA 3GHz \citep{Smolcic_2017}, and MIGHTEE MeerKAT \citep{Heywood_2022}.
Cross matching was performed with a search radius of $\Delta r(\text{subm-mm/radio}) =  1\arcsec$.
More detail on objects overlapping ancillary catalogs can be found in Table \ref{table:champs-overlaps}.
Flags are included in the resulting catalog depicting which, if any, ancillary datasets the CHAMPS sources overlap (see Table \ref{table:catalog-description}).
% Due to the lower resolution at these wavelengths, we used a larger search radius ($\Delta r(\text{subm-mm/radio})\le 1\arcsec$) when identifying matches with these catalogs. 
% We have included flags that indicate which, if any, catalogs the CHAMPS detections may match with (see Table \ref{table:catalog-description}).

% To further validate the $>5 \sigma$ sample, we investigated the 15 $>5 \sigma$ detections not identified when cross matching with COSMOS2025.
% Of these, 2 sources (\textcolor{red}{\textbf{[SOURCE IDs]}}) were just outside our matching radius of $\Delta r(\text{NIRCam})\le 0.6\arcsec$ (but still within $<1\arcsec$) of a cataloged COSMOS2025 object and matched with detections in at least two sub-mm/radio catalogs.
% Due to the proximity of the detections, both sources have since been updated as a match with the COSMOS2025 catalog.

To further validate the SNR $\ge5 \sigma$ sample, we investigated the 15 SNR $\ge5 \sigma$ detections not identified when cross matching with COSMOS2025.
% After visually inspecting the remaining 16 $>5 \sigma$ detections, 
We found 5 detections (CHAMPS-83, CHAMPS-163, CHAMPS-256, CHAMPS-258, and CHAMPS-363) with overlapping NIRCam/MIRI detections that are positioned near a star/diffraction spike in NIRCam, resulting in the source being masked out of the COSMOS2025 catalog and thus missing in our cross match analysis.
All 5 of these sources were also identified in at least two sub-mm/radio catalogs and are therefore verified objects.
Of the remaining 10 SNR $\ge5\sigma$ detections, only 1 (CHAMPS-174) has an SNR above 6 ($\sigma = 7.56$). Unfortunately,
this source falls just off the NIRCam footprint in COSMOS-Web but matches with an object in the A3COSMOS catalog with an offset of $\Delta r (\text{sub-mm})\approx 0.16''$. 
Due to the high SNR and the tight sub-mm detection, we believe this source to be a real SNR $\ge5\sigma$ detection and not a noise fluctuation.

This leaves us with 9 remaining SNR $\ge5 \sigma$ detections that do not overlap prior observations (CHAMPS-327, CHAMPS-348, CHAMPS-349, CHAMPS-358, CHAMPS-362, CHAMPS-365, CHAMPS-369, CHAMPS-373, and CHAMPS-377), all of which are located near an edge of the mosaic with an average SNR $\approx 5.12^{+0.06}_{-0.08}$.
These are likely examples of the noise fluctuations expected to appear along the edges of our mosaic and are within the range of the expected noise peaks we would see with SNRs $\ge5\sigma$ through tests on the completeness (see Section \ref{sec:completeness} for further details).
We have chosen to keep all 15 of these sources in the catalog but have included flags listing which prior catalogs they do (and do not) match with. 
Furthermore, we have excluded the 9 sources that do not overlap prior observations from the analysis of galaxy properties presented in this paper, bringing our working sample of SNR $\ge5\sigma$ sources to 376.

\subsubsection{$4.5\sigma\le$ SNR $<5\sigma$ Blind Detections}
To identify objects that may have been missed with the strict criteria used above, we ran the same peak detection process for sources within the range of $4.5\sigma\le \rm{SNR}<5\sigma$, the lower bound chosen as the SNR where the map noise deviates from a standard normal distribution, highlighted as the black dotted line in Figure \ref{fig:SNR}. 
We find a total of 595 positive peaks, and 146 negative peaks above an SNR $\ge 4.5\sigma$, returning a purity rate of $P(\ge4.5\sigma)=75.46\%$.

With further examination of the sources that are found only in the range of $4.5  - 5 \sigma$, we see 210 positive and 134 negative peaks, returning a local purity rate of $P(4.5\le\sigma<5) = 36.2\%$. 
NIRCam cross matching ($\Delta r(\text{NIRCam})= 1\arcsec$) was performed for the sources within the lower SNR range, returning 120 positive and 34 negative overlaps, resulting in a ``modified purity" rate of $P(4.5\le\sigma<5)_{\rm modified} =71.67\%$.
% A total of 210 positive peakse were identified in the range of $4.5  - 5 \sigma$ while 134 were identified in the negative map, returning a base purity rate of $P(4.5\le\sigma\le5) = 36.2\%$. 
% After performing the same NIRCam cross matching as done for the $>5\sigma$ sample ($\Delta r(\text{NIRCam})= 1\arcsec$), we identified
% 120 positive sources within $4.5 \le \sigma \le 5$, while only matching 34 negative noise peaks, returning a ``modified purity" rate of $P(4.5\le\sigma\le5)_{\rm modified} =71.67\%$.
This confidence is then tightened through the cross examination of sub-mm/radio observations with the search radius $\Delta r(\text{subm-mm/radio})= 1''$. 
For the 90 source candidates with NIRCam detection, 63 match with at least one sub-mm/radio detection, while 39 are detected in at least 2 sub-mm/radio catalogs.

We included the $4.5\le\sigma\le5$ sample in the final blind catalog along with flags depicting which, if any, prior observations they overlap.
While included in the catalog, we do not include the $4.5\le\sigma\le5$ sample in the analysis in this paper in order to ensure the most robust results.
We leave it to the user's discretion to further investigate the $4.5\le\sigma\le5$ objects.

\subsection{MIRI Prior Detections}
\begin{figure}[h]
    \begin{center}
         \includegraphics[trim=0mm 0mm 0mm 0mm, clip, width=1\columnwidth]{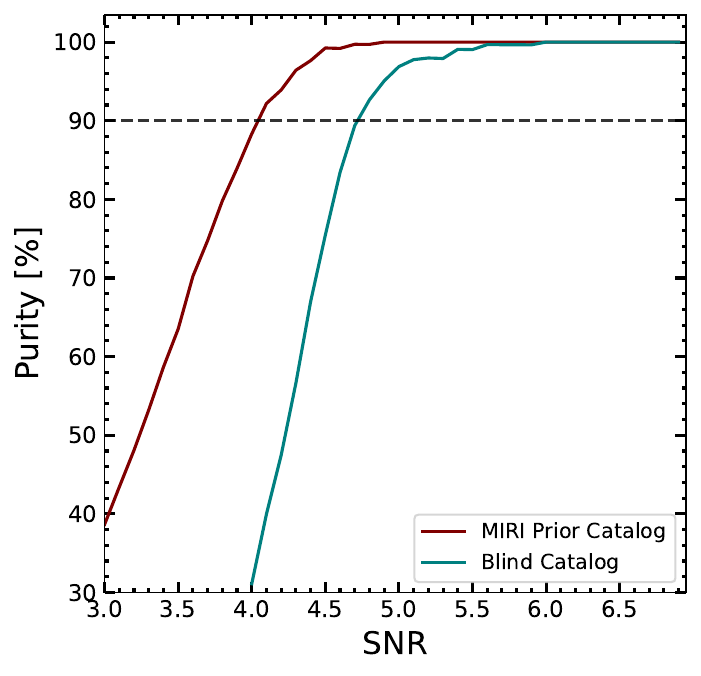}
         \caption{Comparison on the purity estimates of the blind catalog (teal) against the MIRI prior catalog (maroon). We find that the MIRI prior catalog reaches a $90\%$ purity at a lower SNR threshold (SNR $\sim4\sigma$) than the blind catalog (SNR $\sim4.7\sigma$).
         }
         \label{fig:purity-compare}
    \end{center}
\end{figure}

We also created a prior-based catalog using JWST/MIRI detections to identify peaks in the CHAMPS mosaic as a way to reach fainter sources at SNRs $< 5\sigma$.
A CHAMPS source was added to the prior catalog if the CHAMPS SNR map peaked above $\ge3\sigma$ within $1''$ of a COSMOS2025 catalog source with a MIRI detection \citep{Shuntov_2025}.
Like with the blind catalog, if multiple MIRI detections were identified within this range, favorability was given to the source with the reddest color ($m_{\text{F277W}} - m_{\text{F444W}}$).
To ensure we accurately matched the corresponding MIRI source to the correct object identified in CHAMPS, 
a local maximum search was performed to exclude sources that have a portion of their emission fall within $1''$ of the MIRI prior, but have their peak emission outside the $1''$ search radius.
This local maximum test removed objects that may have been incorrectly matched to nearby sources by keeping to a more narrow search window.
This search method was performed on the positive and negative maps returning 3162 positive and 1944 negative MIRI overlapping detections above a peak SNR $\ge 3 \sigma$ in the CHAMPS mosaic, resulting in a purity metric of $P(\ge3\sigma)_{\rm MIRI \ Prior} = 38.52\%$.
Observing sources found with an SNR $\ge4\sigma$ returns a purity of $P(\ge4\sigma)_{\rm MIRI \ Prior} = 88.34\%$ while sources with an SNR $\ge5\sigma$ results in a purity of $P(\ge3\sigma)_{\rm MIRI \ Prior} = 100\%$.
We see that the MIRI prior catalog reaches a $90\%$ purity at a lower SNR threshold than the blind catalog (see Figure \ref{fig:purity-compare}).

% \begin{figure}
%     \begin{center}
%         % \includegraphics[trim=0mm 0mm 0mm 0mm, clip, width=1\columnwidth]{figures/prior_submm_crossmatch_highsnr.pdf}
%         % \includegraphics[trim=0mm 0mm 0mm 0mm, clip, width=1\columnwidth]{figures/prior_submm_crossmatch_lowsnr.pdf}
%          \includegraphics[trim=0mm 0mm 0mm 0mm, clip, width=1\columnwidth]{figures/prior_submm_crossmatch_line.pdf}
%          \caption{The fraction of sources from the MIRI prior catalog that are also observed in ancillary sub-mm/radio observations as a function of SNR.
%          We display the fraction of sources that are not identified in any sub-mm/radio catalog (maroon), the fraction of sources in at least one sub-mm/radio catalog (green), and the fraction of sources in at least two sub-mm/radio catalogs (orange).
%          % We find that below an SNR $<5\sigma$, the fraction of CHAMPS sources that are observed in at least one sub-mm/radio catalog falls to $\sim30\%$ while the reminder of sources are only overlapped by MIRI detections.
%          }
%          \label{fig:submm-crossmatch}
%     \end{center}
% \end{figure}

Further cross matching was performed on the MIRI prior catalog with the sub-mm/radio catalogs populating the COSMOS field.
Following the same methods described in Section \ref{sec:blind_detections}, we identify
Much like the analysis that was performed for the blind $>5\sigma$ sample described in Section \ref{sec:blind_detections}, we listed a source as a match if the source fell within $\Delta r(\text{subm-mm/radio})\le 1''$.
We found 1334 sources identified in at least one sub-mm/radio catalog and 430 sources identified in at least two sub-mm/radio catalogs.
More detail on objects overlapping ancillary catalogs can be found in Table \ref{table:champs-overlaps}.

\begin{figure}[h]
    \begin{center}
         \includegraphics[trim=0mm 0mm 0mm 0mm, clip, width=1\columnwidth]{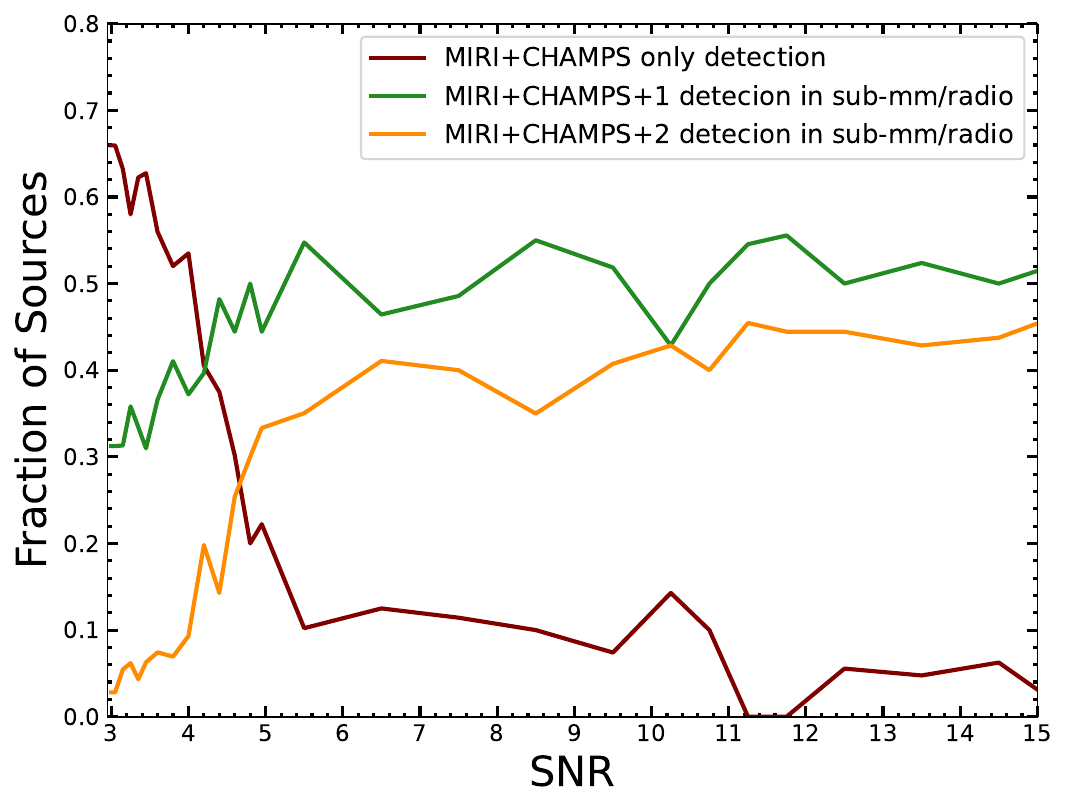}
         \caption{The fraction of sources from the MIRI prior catalog that are also observed in ancillary sub-mm/radio observations as a function of SNR.
         We display the fraction of sources that are not identified in any sub-mm/radio catalog (maroon line), the fraction of sources in at least one sub-mm/radio catalog (green line), and the fraction of sources in at least two sub-mm/radio catalogs (orange line).
         % We find that below an SNR $<5\sigma$, the fraction of CHAMPS sources that are observed in at least one sub-mm/radio catalog falls to $\sim30\%$ while the reminder of sources are only overlapped by MIRI detections.
         }
         \label{fig:submm-crossmatch}
    \end{center}
\end{figure}

Figure \ref{fig:submm-crossmatch} displays the fraction of sources from the MIRI prior catalog that are also observed in ancillary sub-mm/radio detections as a function of SNR.
We found that while the fraction of sources that did not match with any sub-mm/radio catalog increases at SNRs below $< 5 \sigma$, the number of sources that are seen in at least one sub-mm/radio catalog and in at least two sub-mm/radio catalogs stays consistent even down to low SNRs.
This suggests that the MIRI prior sources identified in multiple sub-mm/radio catalogs are likely true detections.
The increasing number of sources that are not seen in any sub-mm/radio catalog for SNRs $\le 5 \sigma$ may come from a combination of faint objects being missed in prior programs and possible noise peaks identified in the mosaic.
Because of this uncertainty, we leave it to the user's discretion to further investigate the objects with SNRs $<5\sigma$ that hold no sub-mm/radio overlap.

\begin{table*}[t]
    \begin{center}
        \includegraphics[trim=0mm 0mm 0mm 0mm, clip, width=2\columnwidth]{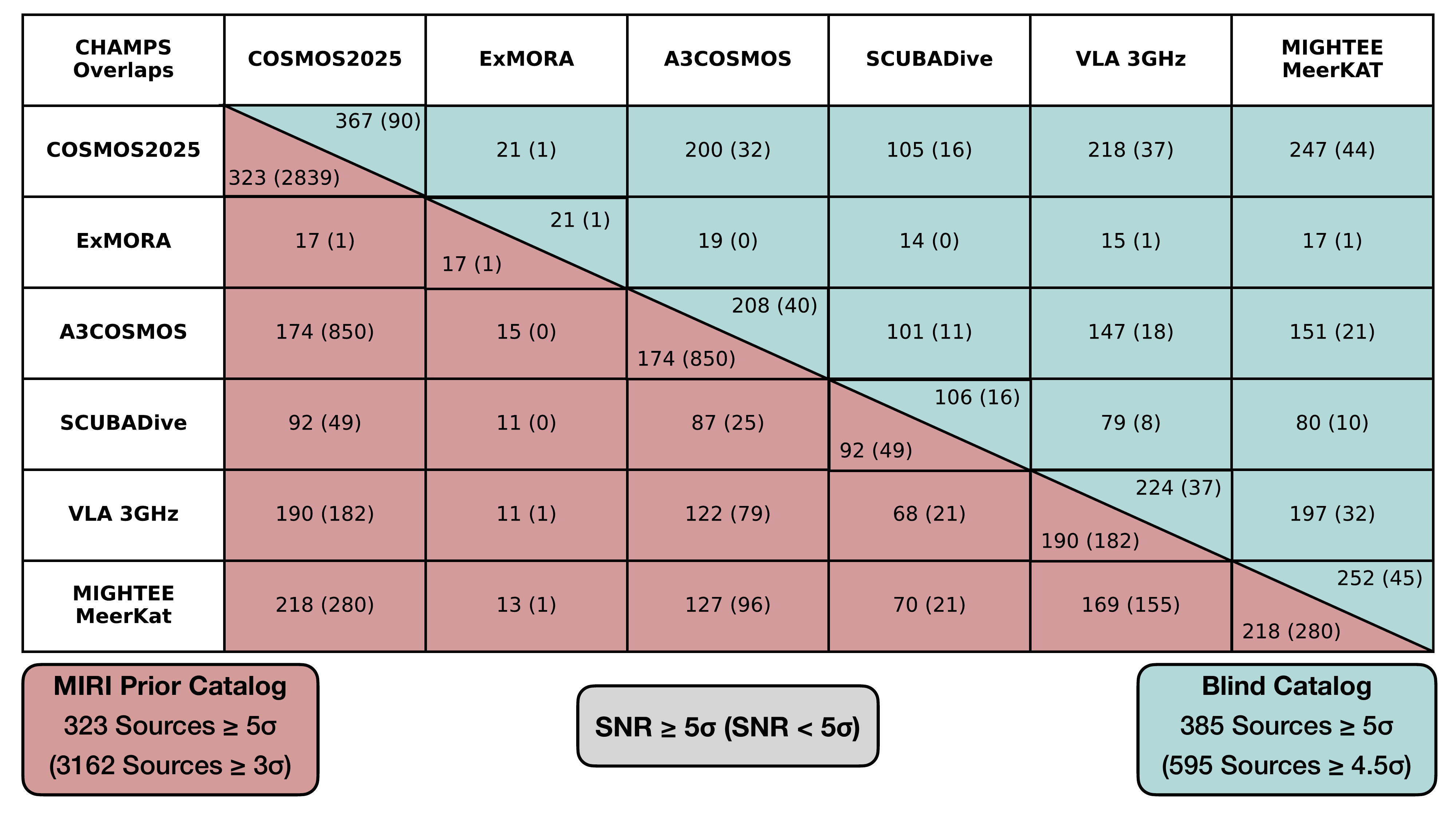}
         \caption{Table showing the number sources in the blind catalog (teal shaded region) and the MIRI prior catalog (maroon shaded region) that were identified within $1\arcsec$ of ancillary programs. As denote by the gray box at the bottom, the first value in each cell represents the number of $\ge5\sigma$ sources that match with the ancillary catalogs while the second value in the parenthesis represents the number of $<5\sigma$ sources that match in the same catalogs.} \label{table:champs-overlaps}
    \end{center}
\end{table*}

\section{Catalog}\label{sec:catalog}

% Visually inspecting these sources found XX neighboring a nearby point source with prior JWST photometry. This likely resulted in the source being masked out when cataloging. Only XX sources are highlighted that have no emission in prior JWST/NIRCam+MIRI photometry, these sources all have peak SNR estimates near $5\sigma$ and are therefore likely noise spikes seen in our mosaic.

% We also hold 46 sources that were identified with our MIRI prior detection method,

\subsection{Creation of the Catalogs}
We compile two catalogs based on the two detection methods described in Section 4. 
A blind detection catalog with sources down to SNR $\ge 4.5\sigma$, and a JWST/MIRI prior detection catalog with sources down to SNR $\ge 3\sigma$.
For the blind catalog, we have identified 595 total sources above SNR $\ge 4.5\sigma$, with 385 identified with an SNR $\ge 5\sigma$, and 210 within an SNR range of $4.5\sigma \le$ SNR $< 5\sigma$. 
% We display the number of sources that were identified in other programs after cross matching, as described in Section \ref{sec:blind_detections}, in Table \ref{table:champs-overlaps}.
For the JWST/MIRI prior catalog, we have identified 3162 sources above SNR $\ge 3\sigma$, where 323 have an SNR $\ge 5\sigma$. 
Between the two catalogs, 402 sources are identified in both the blind and MIRI-prior catalogs leaving a total of 3355 unique objects identified in CHAMPS.
All 323 SNR $\ge 5\sigma$ sources in the MIRI prior catalog are identified in the blind catalog.
A total number of 385 SNR$\ge5\sigma$ sources have been identified, cutouts of which can be seen in Figure \ref{fig:cutouts} in the Appendix.

Numbering for CHAMPS detections is performed in order of SNR where the highest SNR detection (SNR=37.27) is numbered as CHAMPS-1. Sources that are unique to the prior catalog are given an additional +1000 to their identification number resulting in a gap between CHAMPS-595, the last source listed in the blind catalog, and CHAMPS-1001, the first source that is only identified in the MIRI prior catalog.
We show the number of sources that were identified in other programs in Table \ref{table:champs-overlaps} after the same cross matching described in Section \ref{sec:blind_detections}.
Included in both the blind catalog and the JWST/MIRI prior catalog are flags for each source indicating which survey has previously detected the object as well as the survey's corresponding object ID as seen in Table \ref{table:catalog-description}.

Finally, we note that both the blind and MIRI-prior catalogs will be provided upon the acceptance of this paper but can made available early upon request.

\subsection{Flux Measurements}\label{sec:flux}

Peak flux measurements and errors were identified by finding the pixel location of the maximum SNR value for each object and tracing that location onto the primary beam corrected mosaic ($M_\text{PBcorr}$) to obtain the peak flux measurement, as well as onto the RMS map ($M_\text{RMS}$) to obtain the peak flux error for all sources identified in both the blind and prior catalogs.

We use \texttt{PyBDSF} (the Python Blob Detector and Source Finder, \citealp{Mohan_2015}) to obtain the total flux density and errors for all sources in both catalogs.
This code has a targeted functionality that fits 2D Gaussian functions to the blobs that make up a standard sub-mm source.
Fitting is performed on the primary beam corrected mosaic ($M_{\text{PBcorr}}$) with positions obtained from the location of a sources' peak signal identified in the SNR map ($M_\text{SNR}$).
This process returns the total flux density and its corresponding error for every source.

\begin{figure}[h!]
    \begin{center}
        \includegraphics[trim=0mm 0mm 0mm 0mm, clip, width=1\columnwidth]{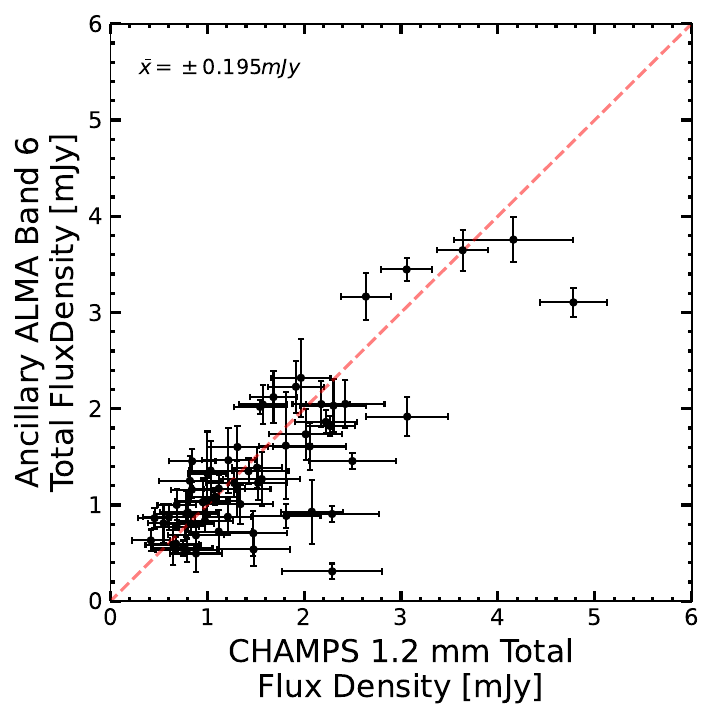}
         \caption{CHAMPS 1.2 mm total flux density plotted against the total flux density of ancillary ALMA band 6 observations (ranging from 1.1 mm - 1.3 mm) with the mean offset displayed ($\bar{x}$). Ancillary data were obtained from A3COSMOS \citep{Liu_2019,Adscheid_2024}. The red dashed line represents the 1-to-1 relation.
         We see that 1.2 mm flux density of CHAMPS sources closely follow the 1-to-1 relation of previous band 6 observations, suggesting that the CHAMPS reduction was performed correctly and that the 1.2 mm total flux density measurements are accurate.} \label{fig:flux_comparison}
    \end{center}
\end{figure}

We compare the total flux density for CHAMPS sources with previous ALMA band 6 observations.
There are 73 $>5 \sigma$ CHAMPS sources that have ancillary ALMA band 6 detections as presented by the cross-matching performed in Section \ref{sec:blind_detections}.
Figure \ref{fig:flux_comparison} shows the relationship between the CHAMPS total flux density compared to that of ancillary data.
We identify a strong 1-to-1 correlation when including errors for a majority of the 73 overlapping sources.
This suggests that the reduction was performed consistent with previous work and that the 1.2 mm total flux density measurements for the catalog are accurate. 
Only a handful of CHAMPS sources were found to have a higher total flux density from previous observations, which could result from a difference in beam sizes and resolutions between CHAMPS and said previous programs.

\subsection{Number Counts}\label{sec:num-density}

\begin{figure}[h!]
    \begin{center}
        \includegraphics[trim=0mm 0mm 0mm 0mm, clip, width=1\columnwidth]{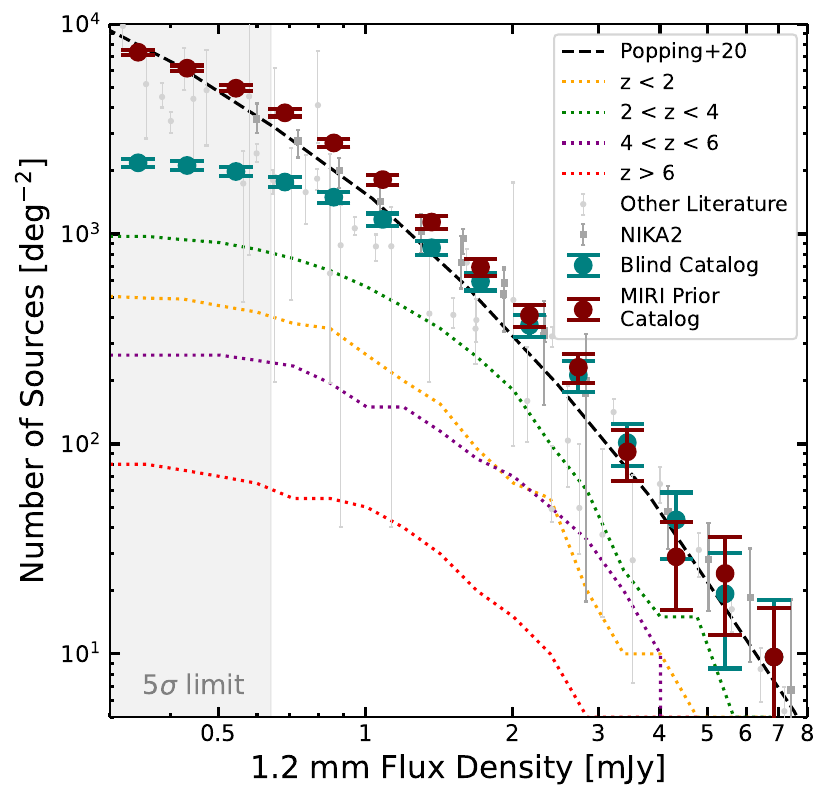}
         \caption{Total 1.2 mm cumulative number density counts observed in the blind catalog (teal) and MIRI prior catalog (maroon).
         Error bars include Poisson and photometric uncertainties. 
         1.2 mm cumulative number density redshift bins are shown as dotted lines ($z<2$ in orange; $2<z<4$ in green; $4<z<6$ in purple; $z>6$ in red).
         The median $5\sigma$ limit is shown in the gray shaded region. 
         Also shown is the model by \citet{Popping_2020} (for all galaxies at $z<7$; dashed line) and measurements from the NIKA2 Cosmological Legacy Survey \citep[N2CLS, dark gray squares;][]{Bing_2023,Bethermin_2026,CarvajalBohorquez_2026}, and other literature (light gray points) including the ASPECS survey \citep{Aravena_2016} and various measurement from the compilation by \citet{Bing_2023} (see text for references).
         } \label{fig:number-density}
    \end{center}
\end{figure}

The cumulative number density is defined as the on-sky surface density of sources above a certain flux density threshold.
The cumulative number density of 1.2 mm sources in both catalogs (prior catalog as blue circles, blind catalog as black circles) is shown in Figure \ref{fig:number-density} and Table \ref{table:number-counts} with errors including Poisson ($\propto\sqrt{N}$) and photometric uncertainties.
The Poisson errors were calculated without approximation using the Python function \texttt{poisson\_conf\_interval()} as part of the \texttt{astropy.stats} package.

The number counts are compared to the model by \citet{Popping_2020} (which includes all galaxies at $z<7$) and measurements from the NIKA2 Cosmological Legacy Survey \citep[N2CLS;][]{Bing_2023,Bethermin_2026,CarvajalBohorquez_2026}, the ASPECS survey \citep{Aravena_2016}, as well as other literature from the compilation by \citet{Bing_2023} \citep[including][]{Lindner_2011,Scott_2012,Fujimoto_2016,Hatsukade_2016,Umehata_2017,Hatsukade_2018,GonzalezLopez_2020,GomezGuijarro_2022,Chen_2023}.
We note that the number counts for the blind catalog start to turn over around 1 mJy, while the prior catalog continues to follow the model. 
% This is a result of the completeness
This likely results from the prior catalog probing down to fainter sources that the blind catalog may withhold.
% which can be seen through the completeness calculations in Figure \ref{fig:completeness}.
% We must also note that the blind catalog detects an increase of spike of $\sim4\,\rm mJy$ objects identified in the blind catalog that a

We further display the 1.2 mm cumulative number density of our sample in redshift bins which are discussed further in Section \ref{sec:redshift}. We see that the $z>6$ bin has the fewest sources detected, while the $2<z<4$ bin around cosmic noon has the most sources detected.

% which is where the survey becomes less complete \citep[for details see][]{Martinez2026}.

% \textcolor{red}{\textbf{[Include ancillary data in Figure 8 and add discussion about how the CHAMPS number density compares to previous estimates]}}

\noindent\begin{table}[h!]
  \caption{CHAMPS 1.2 mm Cumulative Number Densities}
    \begin{tabular}{c|cc}
    
    \midrule\midrule
     & Blind catalog & Prior catalog \\
    \midrule\midrule
    $S_{1.2 \rm mm}$  &  $N(>S_{1.2 \rm mm})$ & $N(>S_{1.2 \rm mm})$ \\
    (mJy)  & (deg$^{-2}$) & (deg$^{-2}$) \\
    \Xhline{1pt}

    0.34 & $2184^{+10}_{-10}$ & $7343^{+63}_{-68}$ \\ 
    0.43 & $2121^{+14}_{-19}$ & $6164^{+63}_{-63}$ \\ 
    0.54 & $1986^{+19}_{-19}$ & $4947^{+63}_{-54}$ \\ 
    0.68 & $1768^{+24}_{-24}$ & $3778^{+58}_{-63}$ \\ 
    0.86 & $1498^{+24}_{-29}$ & $2715^{+58}_{-53}$ \\ 
    1.08 & $1174^{+24}_{-24}$ & $1816^{+43}_{-43}$ \\ 
    1.36 & $860^{+19}_{-24}$ & $1140^{+34}_{-34}$ \\ 
    1.72 & $594^{+19}_{-19}$ & $696^{+24}_{-24}$ \\ 
    2.16 & $367^{+14}_{-14}$ & $411^{+15}_{-19}$ \\ 
    2.72 & $213^{+10}_{-14}$ & $232^{+14}_{-14}$ \\ 
    3.43 & $101^{+10}_{-10}$ & $92^{+10}_{-10}$ \\ 
    4.32 & $43^{+5}_{-5}$ & $29^{+5}_{-0}$ \\ 
    5.43 & $19^{+5}_{-5}$ & $24^{+5}_{-5}$ \\ 
    6.84 & $10^{+0}_{-0}$ & $10^{+0}_{-0}$ \\ 
    \midrule\midrule
    \end{tabular}\label{table:number-counts}

\end{table}

\subsection{False Detection Rates, Flux Boosting, $\&$ Completeness}\label{sec:completeness}

% A series of thorough tests were performed to identify survey completeness, false detection rates, and flux boosting effects.
A series of thorough tests were performed to identify survey completeness, false detection rates, and flux boosting effects.
To quantify the false detection rate, we create mock noise maps that are the same size as the CHAMPS mosaic with Gaussian noise distributed randomly. 
We then convolve this map with the restoring synthesized beam ($\theta_{\rm{res}} \approx 1.1'' \times 1.1''$) and re-normalize the convolved map to a standard normal distribution to represent a pure noise (source-free) map. 
We then count the number of serendipitous peaks at or above $5\sigma$ significance. 
This process was repeated 100 times, where we identified an average of $3^{+3}_{-1}$ false detections in the mock CHAMPS maps, with an average SNR of $5.4^{+0.2}_{{-0.2}}$.

While the mock noise maps have the same pixel size as the CHAMPS mosaic, the real CHAMPS image has more edges than the ideal square-shape of the mock map.
The unique shape of the CHAMPS footprint results in more edges found in the observed reduction leading to more regions of higher noise, low primary beam response, and thus more false positives.
To accommodate this bias, an alternative false positive analysis was performed by examining the negative noise map. 
Originally discussed in Section \ref{sec:blind_detections}, this method involves taking the negative SNR map (multiplied by -1) to identify the total number of negative $>5\sigma$ detections.
This test resulted in 12 false detections, 6 of which could be avoided by setting parameters for a primary beam threshold $> 0.3$ or an RMS threshold to the $90\%$ value of 200$\mu$Jy beam$^{-1}$ (as shown in Figure \ref{fig:RMS}).
The 6 remaining sources have average SNR value of 5.3$^{+0.3}_ {-0.2}$.
This is in good agreement with estimates resulting from the mock noise map reduction, as well as the $>5\sigma$ results from the blind catalog for which we identify 10 likely spurious sources at the same SNR range (see Section \ref{sec:blind_detections}).

Survey completeness is calculated by inserting artificial sources at random positions around the mosaic then using the same peak finding algorithm to recover them. 
Before adding the artificial sources, we remove sources in the mosaic by replacing pixels that hold an SNR$\ge4\sigma$ with random noise matching a standard normal distribution (Gaussian) centered at 0 with a $\sigma=1$ (as shown by the red line representing the background noise in Figure \ref{fig:SNR}).
The artificial sources are then convolved with the restoring synthesized beam  ($\theta_{\rm{res}} \approx 1.1'' \times 1.1''$) before adding into a random position on the CHAMPS mosaic.
% These steps are carried out for a given input and flux size.

\begin{figure*}
    \begin{center}
        \includegraphics[trim=0mm 0mm 0mm 0mm, clip, width=2\columnwidth]{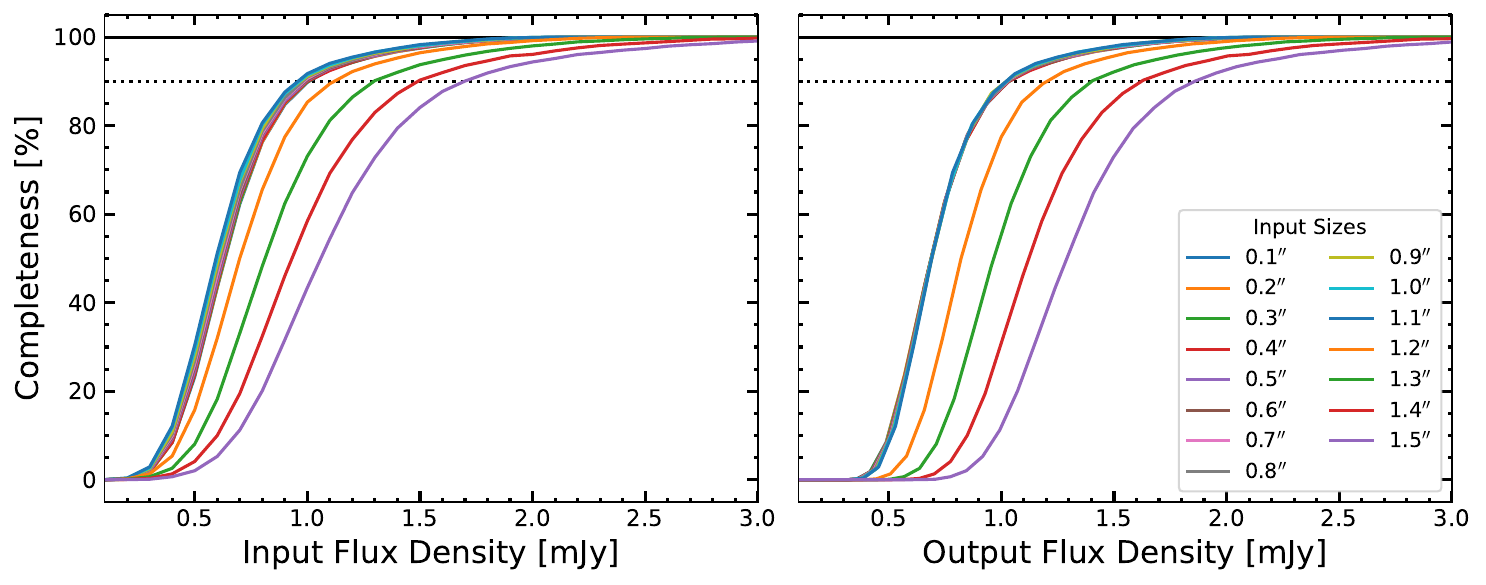}
        \includegraphics[trim=0mm 0mm 0mm 0mm, clip, width=1\columnwidth]{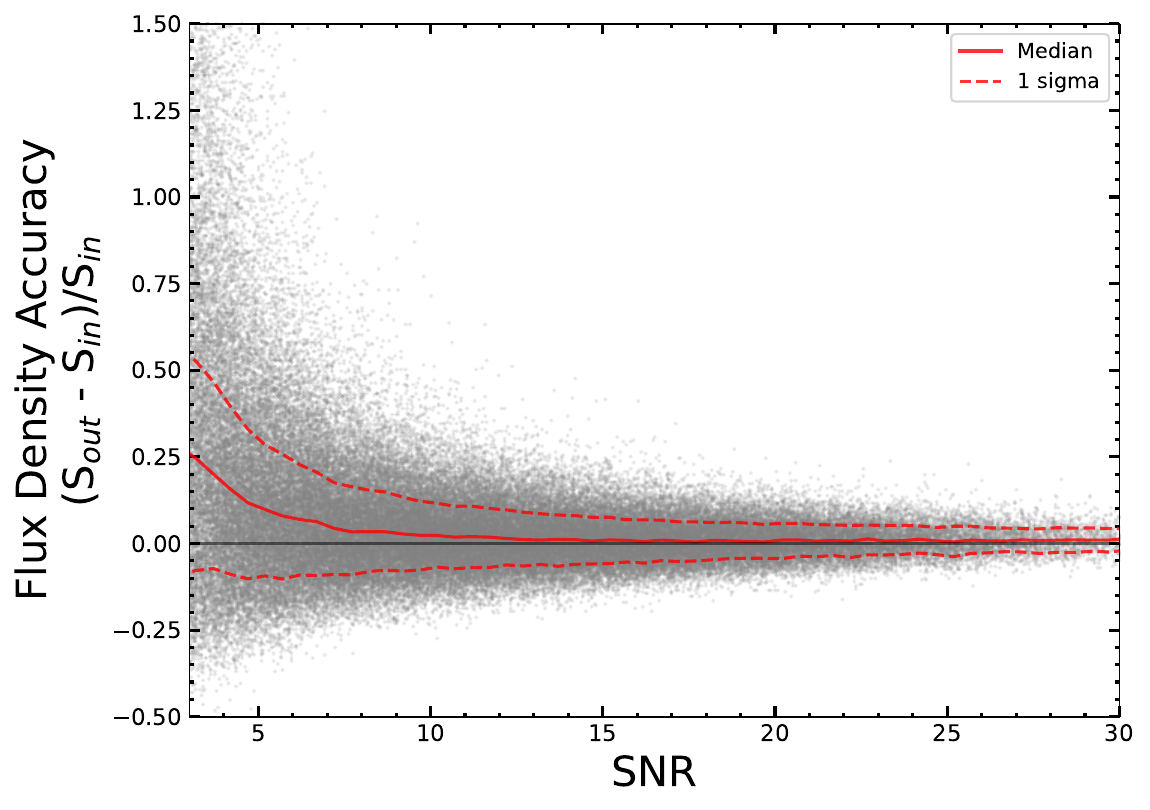}
         \caption{Above, we plot the completeness of CHAMPS as a function of input flux density ($S_{\rm in}$; left) and output flux density ($S_{\rm out}$; right) for different model sources ranging from $0.1\arcsec$ to $1.5\arcsec$ FWHM with their respective colors. 
         The $90\%$ and $100\%$ completeness thresholds are displayed by the horizontal dotted and solid lines respectively.
         Below, we plot the Flux Density Accuracy ($(S_{\rm out} - S_{\rm in})/S_{\rm in}$) as a function of SNR. The solid red line shows the median result with the dashed lines showing $\pm1\sigma$ spread.
         % We see that sources below SNR $\le5\sigma$ experience minor boosting effects.
         We detect a median flux-boosting factor of $\sim10\%$ for sources around SNR $\sim5\sigma$.
         } 
         \label{fig:completeness}
    \end{center}
\end{figure*}

We injected 500 sources for a given input flux and size.
The simulations for completeness were carried out for flux densities ranging from 0.1-3 mJy in steps of 0.1 mJy and sizes from 0.1$''$ to 1.5$''$ full width half max (FWHM) in steps of 0.1$''$ providing a total grid of 30 fluxes and 15 sizes composed of 500 sources each.
This process is repeated 100 times for each given input flux and size, calculating the ratio between recovered and inserted sources on each iteration and taking the average of the results as the injected source's completeness.
We plot the completeness as a function of input flux density ($S_{\rm in}$), output flux density ($S_{\rm out}$) for our injected source sizes (Figure \ref{fig:completeness}-top).
Our results find a completeness of $90\%$ at $\sim1$ mJy and $100\%$ at $\sim1.7$ mJy for both input and output flux densities of sources at or below a FWHM of $1.1\arcsec$.
The completeness progressively decays below this value and is also lower for increasing source sizes at fixed flux densities.

These same simulations were used to estimate flux-boosting effects by examining $S_{\rm in}$ and $S_{\rm out}$.
Flux boosting relates to the detection threshold as it reflects the fact that detectable sources at low SNRs are those located in a noise peak and, thus, their flux measurements are systematically boosted, leading to the observed increase of $S_{\rm out}$ compared to $S_{\rm in}$.
% We measure the flux density accuracy ($(S_{\rm out} - S_{\rm in})/S_{\rm in}$) as a function of SNR and find these effects to be minimal: $\sim25\%$ for sources with SNRs $\le 5\sigma$ (Figure \ref{fig:completeness}-bottom). 
We measure the flux density accuracy ($(S_{\rm out} - S_{\rm in})/S_{\rm in}$) as a function of SNR (Figure \ref{fig:completeness}-bottom).
Our results suggest a median flux-boosting factor of $\sim10\%$ for sources just above our SNR $\ge5\sigma$ sample threshold. 
% yet are much lower than the uncertainties on the measured flux densities at a similar range ($\sim15\%$ for a $\sim5\sigma$ detection). 
As expected, we find that flux boosting effects decrease with increasing SNRs and increase with subsequent lower SNRs, reaching a boosting effect of  $\sim25\%$ for the lowest detectable objects with SNRs of $\sim3\sigma$. 
Boosting corrections are not applied to the reported flux density measurements thus any applied corrections must be performed by the user.

% yt
% and find that on aveage
% , a flux boosting factor
% \textbf{\textcolor{red}{[Completeness results are a work in progress...]}}

\subsection{Catalog Limitations}\label{sec:catalog-limits}

As discussed in Section \ref{sec:blind_detections}, there are 9 $>5\sigma$ detections, with an average SNR value of SNR $\approx 5.12^{+0.06}_{-0.08}$, that do not match to any previous JWST/sub-mm/radio detections.
The number and average SNR of these sources fall around the expected number of false detections from the completeness calculations performed in the previous section.
It is worth reiterating that while these sources are likely false positive detections, we include them in the catalog for posterity. 

When examining the number of detections identified in the blind and the MIRI prior catalogs, we see 62 SNR $>5 \sigma$ objects that are detected in the blind catalog that are missing in the MIRI prior.
Removing the aforementioned 9 potentially spurious detections, as well as the 6 sources identified in Section \ref{sec:blind_detections} that either fall along/near a diffraction spike in JWST imaging, or fall outside the COSMOS-Web footprint, leaves 47 SNR $>5 \sigma$ sources unaccounted for in the MIRI prior catalog.
After visual inspection, the majority of these sources are located inside the gaps/holes that appear around the MIRI footprint.
While CHAMPS was designed to cover the MIRI footprint, several pointings fall along or outside the edges of MIRI pointings (see Figure \ref{fig:observations}) which has resulted in CHAMPS identifying galaxies with no MIRI coverage.
This means that while the prior catalog probes deeper, a robust analysis must, at minimum, include the high quality $\ge 5 \sigma$ objects identified the blind catalog as well.

\section{Galaxy Properties}\label{sec:properties}
We use the wealth of ancillary data available in the COSMOS field to study the physical properties of the $\ge5\sigma$ ALMA detected sources.

\subsection{Redshift Distribution}\label{sec:redshift}

We use the photometric redshifts from the COSMOS2025 catalog \citep{Shuntov_2025} that were derived via the use of the template fitting code \texttt{LePHARE} \citep{Arnouts_2002,Ilbert_2006}. 
This results in a photometric redshift measurment for every CHAMPS source that is identified in the COSMOS2025 catalog,
leaving a total of 367 $\ge5\sigma$ sources that hold a photometric redshift (see Table \ref{table:champs-overlaps}). 

% Spectroscopic redshifts were pulled from various sources, notably, the large COSMOS Spectroscopic Redshift Compilation performed by \citealt{Khostovan_2026} (including \citealt{Lilly_2007, Trump_2007, Coil_2011, Casey_2012, Krogager_2014, Scoville_2015, Casey_2015,Kartaltepe_2015,Kriek_2015,Onodera_2015, Onodera_2016, Nanayakkara_2016, Casey_2017, Hasinger_2018, Jin_2019, Kashino_2019,Masters_2019,Wisnioski_2019, Shah_2020, Polletta_2021,vanDerWel_2021, Chen_2022,Jin_2022,Lilly_2023, Cooper_2023, Cooper_2024, Epinat_2024, Gentile_2024,Sillassen_2024, Forrest_2025, AbdulKarim_2026, Lertprasertpong_in_prep, Vanderhoof_in_prep}).

Spectroscopic redshifts were pulled from various sources as described in Section \ref{sec:ancillary}, notably the COSMOS Spectroscopic Redshift Compilation \citep[][]{Khostovan_2026}, DJA \citep{de_Graaff_2024,Heintz_2025,Valentino_2025,Gillman_2026}, and COSMOS-3D  (PID $\#$5893, PI: K. Kakiichi, \citealt{Koki_2024}).

Various other NIRSpec/MSA observations were used for identifying spectroscopic redshifts, most notably from CAPERS (GO$\#$6368, PI: M. Dickinson), Mirage or Miracle? (MOM; GO$\#$5224, PI: P. Oesch), EMBER (GO$\#$7076, PI: H. Akins), and ZENITH (GO$\#$7417, PI: C. Casey). 
We reduce all NIRSpec data using a custom wrapper around the JWST Science Calibration Pipeline v1.20.0 using Calibration Reference Data System (CRDS) mapping \texttt{pmap-1481}. 
We largely adopt the default \texttt{jwst} pipeline parameters, but implement some custom routines to better remove $1/f$ noise and the thermal ``picture frame'' effect, as well as \citet{Horne_1986} optimal extraction of 1D spectra. 
Redshifts are fit automatically to all spectra using a non-negative least squares template fitter with templates for common strong emission lines and galaxy/AGN continua. 
All redshifts are visually inspected and assigned a quality ranking based on their reliability (i.e., single or multiple emission lines anchoring the redshift). 

We have identified 148 $\ge5\sigma$ sources with a spectroscopic redshift measurement, of which 70 have a quality flag ($Q_f$) $\ge3$, which is considered as a high-confidence result with a low probability of an incorrect identification. 
Overall, CHAMPS has identified 375 $>5\sigma$ ALMA sources, 68 of which contain both a photometric redshift and spectroscopic redshift above a $Q_f \ge3$, while 299 have only a photometric redshift measurement, and 2 hold only a spectroscopic redshift above a $Q_f >3$, bringing our total number of $5\sigma$ sources with a redshift value to 369.

\begin{figure}[h!]
    \begin{center}
        \includegraphics[trim=0mm 0mm 0mm 0mm, clip, width=1\columnwidth]{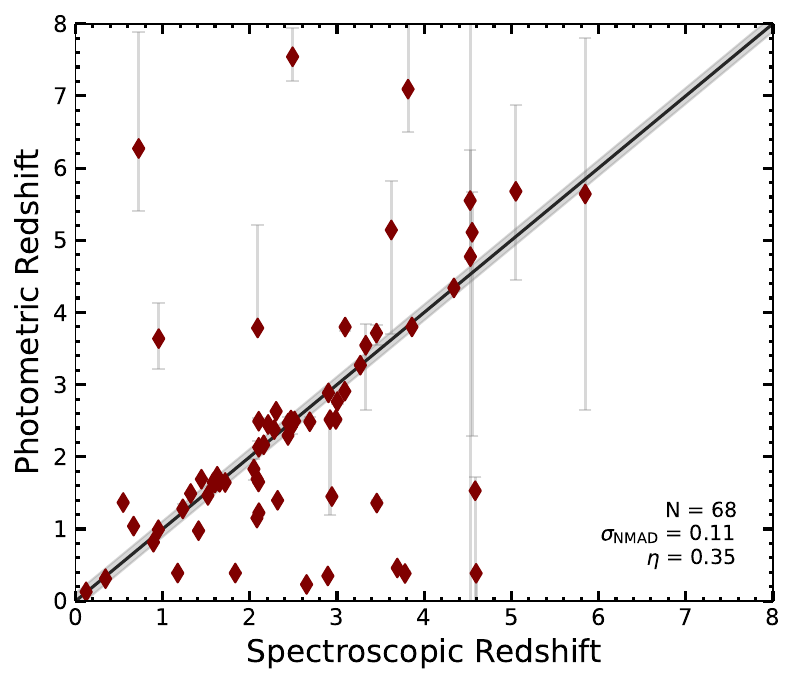}
         \caption{Photometric vs spectroscopic redshift values for $>5\sigma$ ALMA detections with a spectroscopic quality factor $\ge3$. Solid black line represents the one-to-one relation, the shaded gray region represents the normalized median absolute deviation ($\sigma_{\rm NMAD}$) for the sample, and the outlier fraction $(f_\text{out})$ is reported in the bottom right.} \label{fig:redshift-comp}
    \end{center}
\end{figure}

Figure \ref{fig:redshift-comp} compares the photometric and spectroscopic redshift measurements for the 68 sources that contain both in order to assess the performance of the photometric redshifts for this sample of dusty galaxies.
% through a comparison of high-quality flag spectroscopic redshifts.
We quantify this comparison using the normalized median absolute deviation ($\sigma_{\rm NMAD}$) and outlier fraction ($\eta$). NMAD can be calculated as:
\begin{equation}
    \sigma_{\rm NMAD} = 1.48 \times \text{median}\Big(\frac{\Delta z - \text{median}(\Delta z)}{1 + z_{\rm spec}}\Big),
\end{equation}
where $\Delta z$ is defined as the difference between $z_{\rm phot} - z_{\rm spec}$. For the outlier fraction, $\eta$, we consider a source to be an outlier when $|\Delta z| > 0.15 \ (1 + z_{\rm spec})$ as defined by \citealt{Hildebrandt_2012}.
We measure an NMAD score of $\sigma_{\rm NMAD}=0.11$ and an outlier fraction of $\eta = 0.35$, a result suggesting that redshift estimates for dusty galaxies may be difficult to determine photometrically.
% This high fraction of catastrophic outliers suggests that photometric derive redshift estimates may not be suit
% We see that these dusty galaxies present a unique challenge to fitting photometrically.
% This high result could be that the photometric redshifts were not fit with ALMA CHAMPS photometry...

We identified 19 sources with a photometric redshift estimate above $z_{\rm phot} > 6$ that hold no spectroscopic redshift measurment, 3 of which are estimated to have a $z_{\rm phot} > 8$.
However interesting, we note that all 3 sources in the 5$\sigma$ catalog that hold both a photometric redshift estimate above $z_{\rm phot} > 6$ and a high-quality spectroscopic redshift estimate has resulted in the spectroscopic redshift to correct downwards to a lower redshift solution as seen in Figure \ref{fig:redshift-comp}. 
This is an unfortunate characteristic of photometric redshift estimates of DSFGs, as their heavy dust obscuration increases the uncertainty in their photometric redshifts \citep{Zavala_2023,McKinney_2025, Long_2026, McKay_2026}.

\begin{figure*}[t]
    \begin{center}
        \includegraphics[trim=0mm 0mm 0mm 0mm, clip, width=2\columnwidth]{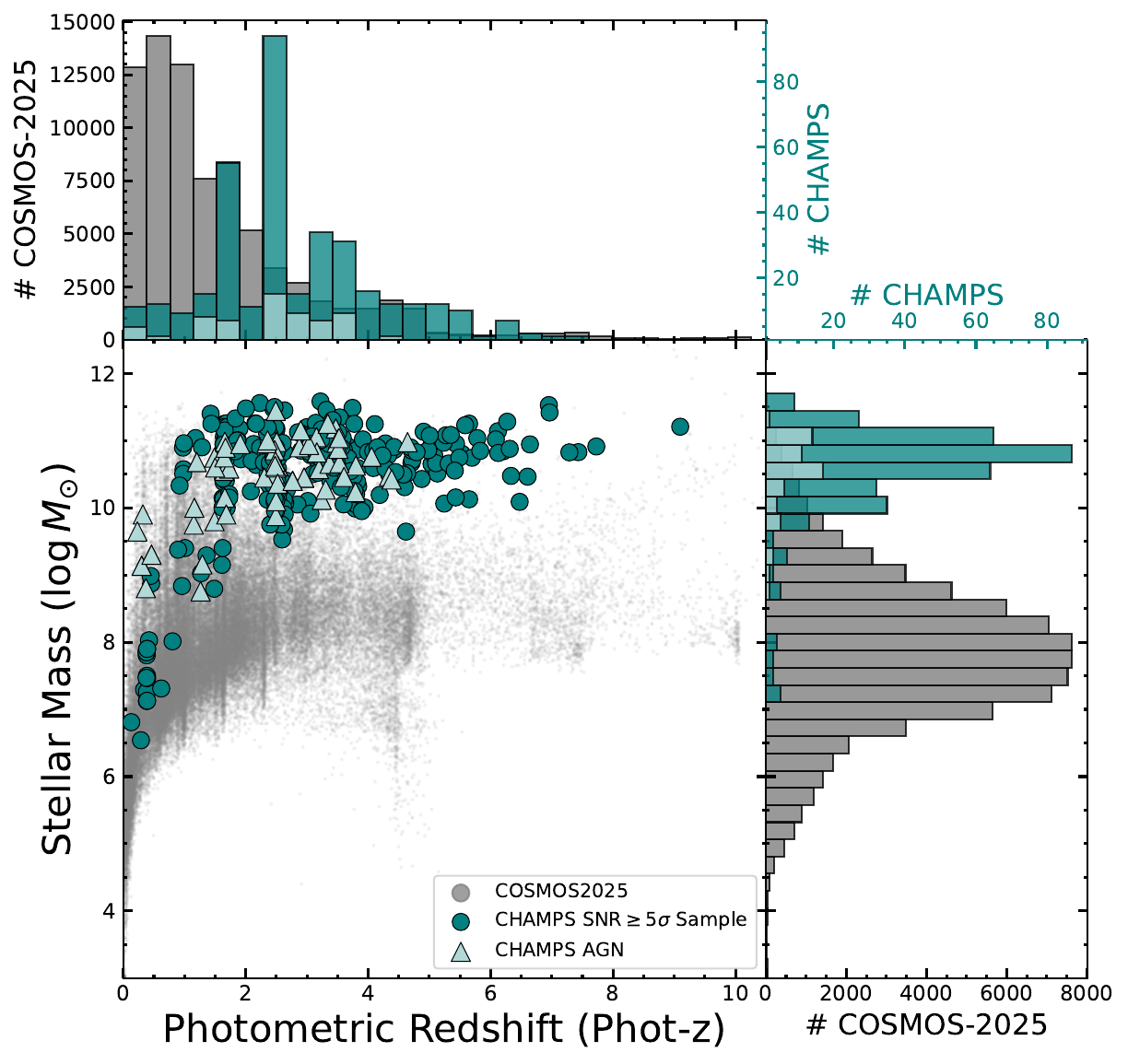}
         \caption{Stellar mass plotted as a function of photometric redshift for 5$\sigma$ sources identified in CHAMPS (teal circles) compared to the standard population of galaxies identified in the COSMOS2025 catalog \citep[small gray dots;][]{Shuntov_2025}. 
         CHAMPS AGN candidates are included (pale teal triangles). 
         The upper histograms shows the photometric redshift distribution of the COSMOS2025 catalog (gray histogram) compared to a scaled-up version of CHAMPS $\ge5\sigma$ sources with the AGN sample inset. 
         The right histogram shows the stellar mass distribution of the COSMOS2025 catalog (gray histogram) compared to a scaled-up version of CHAMPS $\ge5\sigma$ sources with the AGN sample inset. 
         We observe that CHAMPS identifies a population of massive sources one order of magnitude (on average) at higher redshifts compared to the COSMOS2025 catalog.
         } \label{fig:redshift-mass}
    \end{center}
\end{figure*}

\subsection{Stellar Masses}\label{sec:mass}

We also use the stellar mass estimates for the CHAMPS detections from the COSMOS2025 catalog \citep{Shuntov_2025}. 
We find a substantial number of sources in our sample have a stellar mass greater than $10^{10} \rm M_\odot$ (293/367) with a median mass of $\langle\log (\rm M_\star/M_\odot)\rangle = 10.79^{+0.42}_{-3.08}$. 

Figure \ref{fig:redshift-mass} shows the stellar mass as a function of redshift for all $\ge5\sigma$ CHAMPS galaxies with a match in the COSMOS2025 catalog, highlighting that they are more massive than typical SFGs detected in COSMOS-Web at similar redshifts.
The high proportion of massive galaxies among the ALMA detected sources suggest that the depth of CHAMPS is sensitive to the most massive sources at redshifts $z>3$.
% suggests that stellar mass can be a strong driver for a source to be detected by ALMA at high redshift \citep{Dunlop_2017}.
The strong link between detection and stellar mass, paired with the underlying relation between stellar mass and SFR of DSFGs, is consistent with CHAMPS preferentially selecting massive, actively star-forming dusty galaxies.
% suggests that CHAMPS is detecting the most highly star-forming systems at these epochs.
The position of CHAMPS sources along the star formation main sequence, as well as updated SED fits using ALMA data, will be studied in more detail in a forthcoming paper (F. Martinez III et al. in prep).
% The specific star-formation rate (sSFR), defined as the ratio of galaxy SFR to stellar mass, drops quickly at $z<2$, and increases continuously at greater redshifts \citep{Schreiber_2015}.

\subsection{AGN Candidates}\label{sec:agn}

We have identified 68 candidate AGN in the CHAMPS SNR $\ge5\sigma$ sample. 
We used the AGN sample from \citealt{Kelly_prep}, selected using five catalogs covering the COSMOS field: COSMOS2020 \citep{Weaver_2022},
% and COSMOS2025 \citep{Shuntov_2025}, 
the \textit{Chandra} COSMOS Legacy Survey \citep{Elvis_2009,Civano_2016}, the VLA 3 GHz Large Program \citep{Smolcic_2017}, the COSMOS Spectroscopic Redshift Compilation \citep{Khostovan_2026}, and the work released from Dark Energy Spectroscopic Instrument (DESI) DR1 with DESI-COSMOS \citep{Myers_2023,DESI_Collaboration_2026,Ratajczak_2026}.
We summarize the selection methods as follows:

The first selection method utilizes X-ray emission from the \textit{Chandra} COSMOS Legacy Survey.
Sources were flagged as AGN candidates if their intrinsic 0.5-10 keV X-ray luminosity follows $L_{X(0.5-10 \text{ keV})} > 10^{42}$ erg s$^{-1}$.

The second selection uses emission line diagnostics in the rest-frame optical spectra.
This includes the identification of broad emission lines, a common feature found in many AGN spectra resulting from Doppler line broadening occurring around the SMBH, and other emission line diagnostics which were identified with the DESI DR1 release \citep{Myers_2023,DESI_Collaboration_2026}.
\citealt{Kelly_prep}'s analysis uses the COSMOS Spectroscopic Redshift Compilation \citep{Khostovan_2026} and DESI spectroscopic catalogs to identify broadening and other various spectral features that are used to classify AGN.

The third selection method takes advantage of IR emission observed with all four \textit{Spitzer}/IRAC channels to perform a color selection and identify sources with power-law emission in the mid-IR.
% when identifying AGN candidates.
Using the COSMOS2020 \citep{Weaver_2022} catalog, \citealt{Kelly_prep} identifies an AGN at $z<3$ if its IRAC fluxes are monotonically rising $(f_{8.0\  \mu\rm m} > f_{5.8\  \mu\rm m} > f_{4.5\  \mu\rm m} > f_{3.6\  \mu\rm m})$ and whose colors are within the observed-frame color-selection wedge of \citealt{Donley_2012}.

Finally, the fourth selection method utilized excess radio emission to identify AGN candidates. 
\citealt{Kelly_prep} uses the VLA 3 GHz Catalog \citep{Smolcic_2017} and the selection technique described by \citealt{Delvecchio_2017} that defines a radio AGN if the ratio of its 1.4 GHz luminosity ($L_{1.4 \text{ GHz}}$) to the IR-derived star formation rate (SFR$_{\rm IR}$) is more than expected from star formation alone:
\begin{equation}
    \log \Big( \frac{L_{1.4 \text{ GHz}} }{\text{SFR}_{\text{IR}}} \Big)_\text{excess} = 21.984\times(1+z)^{0.013}
\end{equation}
Where SFR$_{\rm IR}$ values are pulled from SED fitting performed in \citealt{Delvecchio_2017}. 

Utilizing multiwavelength selection techniques to identify AGN candidates is good practice as each selection technique has it's own limitations on what types of AGN it can identify, combining multiple techniques can ensure the most robust sample of AGN possible.
\citealt{Kelly_prep}'s catalog performs this process for AGN in COSMOS up to $z\sim12$ for photometrically derived redshifts, and up to $z\sim9$ for spectroscopically confirmed redshifts with the most complete sample at $z<3$ due to the IRAC selection techniques.

Combining these selection criteria with the CHAMPS dataset results in 68/376 AGN candidates, which are highlighted as pale teal triangles in Figure \ref{fig:redshift-mass}.
Of these sources, 19 were identified through their X-ray luminosity, 15 were identified through their spectral line features (including the 4 CHAMPS objects which are identified in \citealt{Zou_2026}: CHAMPS-218, CHAMPS-252, CHAMPS-296, CHAMPS-385),
% (11 through broad line emission, 9 through other QSO spectral features seen in DESI spectroscopy), 
29 candidates were selected with their IR emission via the IRAC color selection, and 26 AGN candidates were identified from their excess radio emission.

We detect the fraction of AGNs of CHAMPS $\ge5\sigma$ sources to be $\sim18\%$.
% , compared to the fraction of AGNs in the total galaxy population $\sim1-2\%$ \citep{}.
This result matches with the previous estimate of $17(^{+16}_{-6})\%$ performed by \citealt{Wang_2013}, work that is of particular note as its sample was obtained from $\sim100$ ALMA-confirmed 870 $\mu$m-selected sources in \textit{Chandra} Deep Field South with unambiguous counterparts, enabling effectively more precise measurements in both X-ray and the far-IR \citep{Casey_2014}.

% A substantial fraction of AGN among the galaxies detected by ALMA may reflect that these sources are likely experiencing a starburst, possibly triggered by a merger that may dramatically reduce the angular momentum of the gas and drive it towards the center of the galaxy \citep[e.g.,][]{Rovilos_2012, Gatti_2015, Lamastra_2013} or cause violent disk instabilities \citep{Bournaud_2012}.
% Alternatively, ALMA might preferentially detect galaxies with a high gas and dust content more prone to efficiently fuel the central black hole and trigger an AGN.

\section{Summary}\label{sec:summary}

In this paper, we present the data reduction and source catalogs of CHAMPS, a new ALMA band 6 (1.2 mm; 250 GHz) blank-field survey covering $0.2$ deg$^2$ of the JWST/NIRCam + MIRI footprint in the COSMOS field.
We find that the average RMS for the CHAMPS mosaic is measured at $146$ $\mu$Jy beam$^{-1}$ and that the majority of the CHAMPS survey (90$\%$) holds an RMS at or below 200 $\mu$Jy beam$^{-1}$ for depths of $5\sigma$ (Figure \ref{fig:RMS}).

CHAMPS identifies sources via two detection methods, a blind search to 
detect high signal sources that unfortunately suffers from a high false-positive detection rate when approaching lower SNRs, and a prior based search involving MIRI detections to identify faint sources that could otherwise have been mistaken for noise with a blind search.
This identification methodology results in two catalogs, a blind catalog, and a MIRI-prior catalog.
% but could potentially bias the sample.
% This results in two ca
% identify faint sources resulting in two catalogs
% high signal sources, and a prior-based search involving MIRI detections to identify faint sources resulting in two catalogs.
The blind catalog holds detections down to SNR $\ge4.5\sigma$ (595 sources total, 385 above SNR $\ge5\sigma$ with a purity $P (\ge5\sigma)_{\rm Blind} = 96.9\%$), while the prior catalog holds detections down to SNR $\ge3\sigma$ (3162 sources total, 323 above SNR $\ge5\sigma$ with a purity $P (\ge5\sigma)_{\rm Prior} = 100\%$).
We have verified 376 $5\sigma$ sources through cross matching ancillary datasets that were used in our analysis of galaxy properties.

Examining the galaxy properties of the 376 SNR $\ge5\sigma$ sources, 68 contain both a photometric redshift and spectroscopic redshift at a $Q_f \ge3$, while 299 have only a photometric redshift measurement, and 2 have only a spectroscopic redshift at a $Q_f \ge3$, bringing the total number of SNR $\ge5\sigma$ sources with a redshift value to 369 and returning a median photometric redshift value of  $\langle z_{1.2 \text{ mm}}\rangle = 2.54^{+0.89}_{-1.69}$.
We found that (293/376) of the SNR $\ge5\sigma$ sources hold a stellar mass greater than $10^{10} \rm M_\odot$ with a median mass of $\langle\log (\rm M_\star/M_\odot)\rangle = 10.79^{+0.42}_{-3.08}$.
Finally, we identify $\sim18\%$ (68/376) of CHAMPS SNR $\ge5\sigma$ sources to be AGN candidates through various selection criteria.

\begin{table*}\
  \caption{CHAMPS catalog column descriptions. Both blind and MIRI-prior catalogs contain the following layout.}
    \begin{tabular}{lll}
    
    \midrule\midrule
    Column Name  & \ \ \ \ \ \ \ \ \ \ \ \ \ \ \ \ \ \ \ \ \  & Description \\
    \Xhline{1pt}
    
    id & &   CHAMPS object ID, ordered by SNR \\
    ra & &   Right ascension [deg] \\
    dec & &   Declination [deg] \\
    peak$\_$flux & &   Peak flux [mJy/beam] \\
    rms & &   RMS  [mJy/beam] \\
    snr & &   SNR  \\
    total$\_$flux & &   Total flux value [mJy] \\
    total$\_$flux$\_$err & &  Total flux error [mJy] \\
    zphot & & Photometric redshift, same as median of the zPDF from COSMOS2025 catalog \\
    zphot$\_$lerr & & Photometric redshift lower $1\sigma$ confidence value, computed from the zPDF \\
    zphot$\_$uerr & & Photometric redshift upper $1\sigma$ confidence value, computed from the zPDF \\
    zspec & & Spectroscopic redshift \\
    zspec$\_$qf & & Quality flag of spectroscopic redshift \\
    zspec$\_$program & & Program spectroscopic redshift was observed in \\
    in$\_$cosmos2025 & &  Has counterpart in COSMOS2025 catalog [boolean] \\
    cosmos2025$\_$id & & COSMOS2025 ID of CHAMPS source \\
    in$\_$a3cosmos & &  Has counterpart in A3COSMOS catalog [boolean] \\
    a3cosmos$\_$id & & A3COSMOS ID of CHAMPS source \\
    in$\_$scubadive & &  Has counterpart in SCUBADive catalog [boolean] \\
    scubadive$\_$id & & SCUBADive ID of CHAMPS source \\
    in$\_$vla3ghz & &  Has counterpart in VLA 3 GHz catalog [boolean] \\
    vla3ghz$\_$id & & VLA 3 GHz ID of CHAMPS source \\
    in$\_$exmora & &  Has counterpart in ExMORA catalog [boolean] \\
    exmora$\_$id & & ExMORA ID of CHAMPS source \\
    in$\_$meerkat & &  Has counterpart in MIGHTEE MeerKAT catalog [boolean] \\
    meerkat$\_$id & & MIGHTEE MeerKAT ID of CHAMPS source \\
    
    \midrule\midrule
    \end{tabular}\label{table:catalog-description}
\end{table*}

\section*{Acknowledgments}\label{sec:ascknowledge}

FM is thankful to Madi Benzor-Martinez for the love and support provided throughout the writing of this manuscript.
The authors are grateful to George Privon and the NRAO computing team for their assistance with running the data reduction and portions of the data analysis on the NRAO-designated computing resources.
JK and FM also thank NRAO for hosting them during the initial phases of the data reduction and for providing valuable guidance on the data reduction procedures. 
FM acknowledges support for this work that was provided by the NSF through award SOSP 1519126 from the NRAO.
The National Radio Astronomy Observatory and Green Bank Observatory are facilities of the U.S.
National Science Foundation operated under cooperative agreement by Associated Universities, Inc. 
This paper makes use of the following ALMA data: ADS/JAO.ALMA$\#$2023.1.00180.L. ALMA is a partnership of ESO (representing its member states), NSF (USA) and NINS (Japan), together with NRC (Canada), MOST and ASIAA (Taiwan), and KASI (Republic of Korea), in cooperation with the Republic of Chile. The Joint ALMA Observatory is operated by ESO, AUI/NRAO and NAOJ.
ET acknowledges support from the ANID CATA-BASAL program FB210003, and FONDECYT Regular 1241005 and 1250821.
HSBA gratefully acknowledges support from Academia Sinica through grant AS-PD-1141-M01-2.
The Cosmic Dawn Center (DAWN) is funded by the Danish National Research Foundation (DNRF140).
Support for this work was provided by the NSF through award SOSP 1519126 from the NRAO. 
TK acknowledges support by the National Science Foundation (NSF) grant No. 2139292.
RAM acknowledges support from the Swiss National Science Foundation (SNSF) through project grant 200020\_207349.
DBS gratefully acknowledges support from NSF Grant 2407752.

\begin{contribution}
%%This section gives authors the space to recognize author contributions. The text inside this environment is NOT counted towards the total word quanta. At a minimum, manuscripts are expected to include this text:

FM is responsible for writing and submitting this manuscript. 
JK, ALF, ASL, JZ, CMC, and JD contributed significantly to the reduction, and source detection results that were presented in this manuscript.
ALF is the PI of CHAMPS. MA, CC, JK, JS, ST, ET, and JZ are co-PIs of CHAMPS and significantly contributed to the survey. 
All other authors contributed by providing comments to this manuscript.
\end{contribution}

\facilities{ALMA}

\software{
\texttt{astropy} \citep{astropy13,astropy18,astropy22}, 
\texttt{CASA} version 6.6.1.17 \citep{Hunter_2023}, \texttt{PyBDSF} \citep{Mohan_2015}, \texttt{photutils} version 1.3.0 \citep{Bradley_2026}
}

%%% FIGURE: INDIVIDUAL GALAXIES %%%%
\begin{figure*}[b!]
\centering
\includegraphics[width=\textwidth,height=0.9\textheight,keepaspectratio]{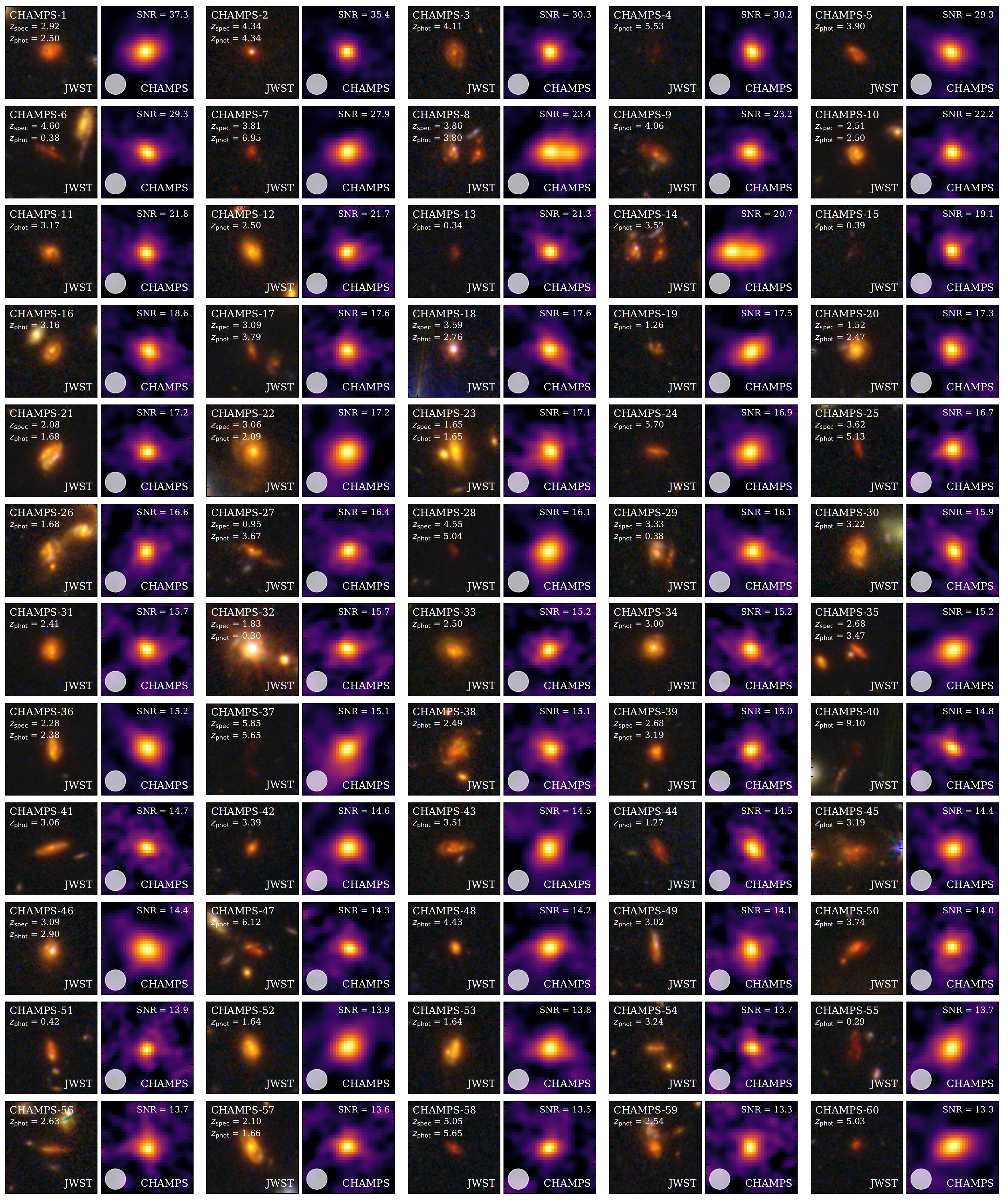}
\caption{
% JWST RGB and CHAMPS 1.2mm cutouts for all CHAMPS SNR $\ge5\sigma$ sources sorted by SNR. 
Cutouts for all CHAMPS SNR $\ge5\sigma$ sources sorted by SNR. 
JWST RGB images (F444W-F277W-F150W) are shown on the left along with photometric and spectroscopic redshift measurements (if available). CHAMPS $1.2\,$mm cutouts are shown on the right with their associated SNR. The cutouts are $5\arcsec$ in size.
\label{fig:collage1}}
\end{figure*}
%%%%%%%%%%%%%%%%%%%%%%%%%%%%%%%

%%% FIGURE: INDIVIDUAL GALAXIES %%%%
\begin{figure*}[b!]
\centering
\includegraphics[width=\textwidth,height=0.9\textheight,keepaspectratio]{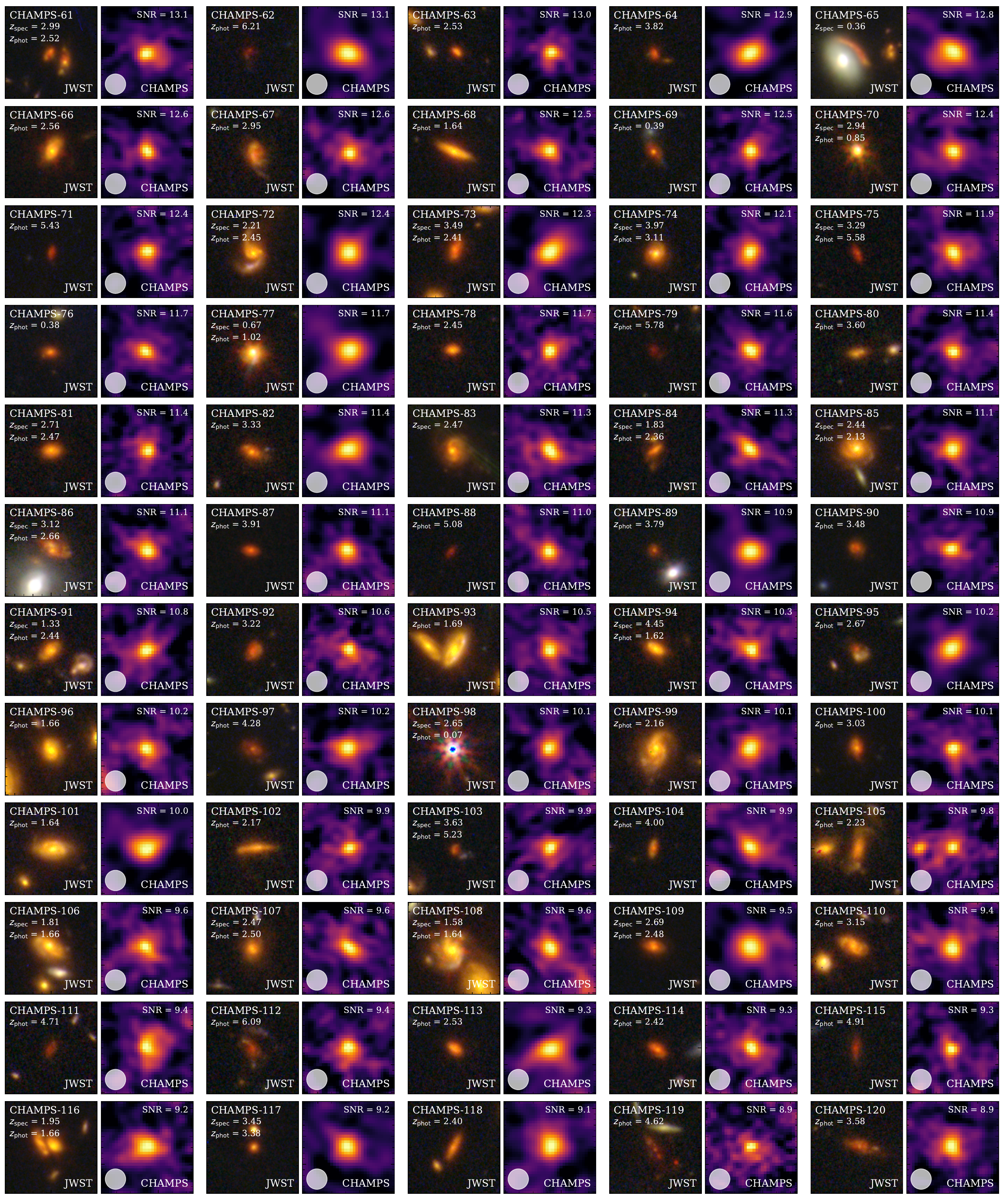}
\caption{Same as Figure~\ref{fig:collage1}.
\label{fig:collage2}}
\end{figure*}
%%%%%%%%%%%%%%%%%%%%%%%%%%%%%%%

%%% FIGURE: INDIVIDUAL GALAXIES %%%%
\begin{figure*}[b!]
\centering
\includegraphics[width=\textwidth,height=0.9\textheight,keepaspectratio]{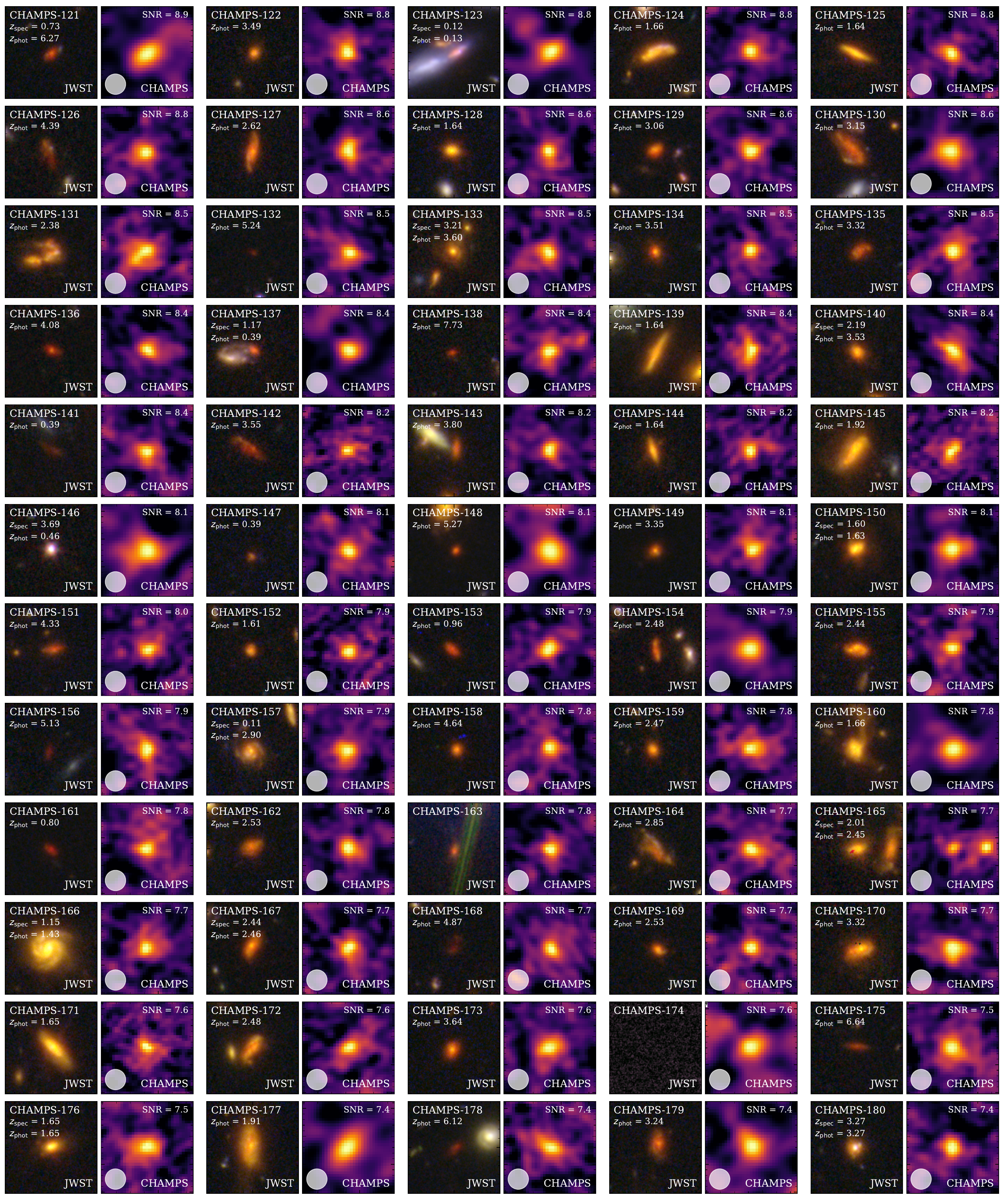}
\caption{Same as Figure~\ref{fig:collage1}.
\label{fig:collage3}}
\end{figure*}
%%%%%%%%%%%%%%%%%%%%%%%%%%%%%%%

%%% FIGURE: INDIVIDUAL GALAXIES %%%%
\begin{figure*}[b!]
\centering
\includegraphics[width=\textwidth,height=0.9\textheight,keepaspectratio]{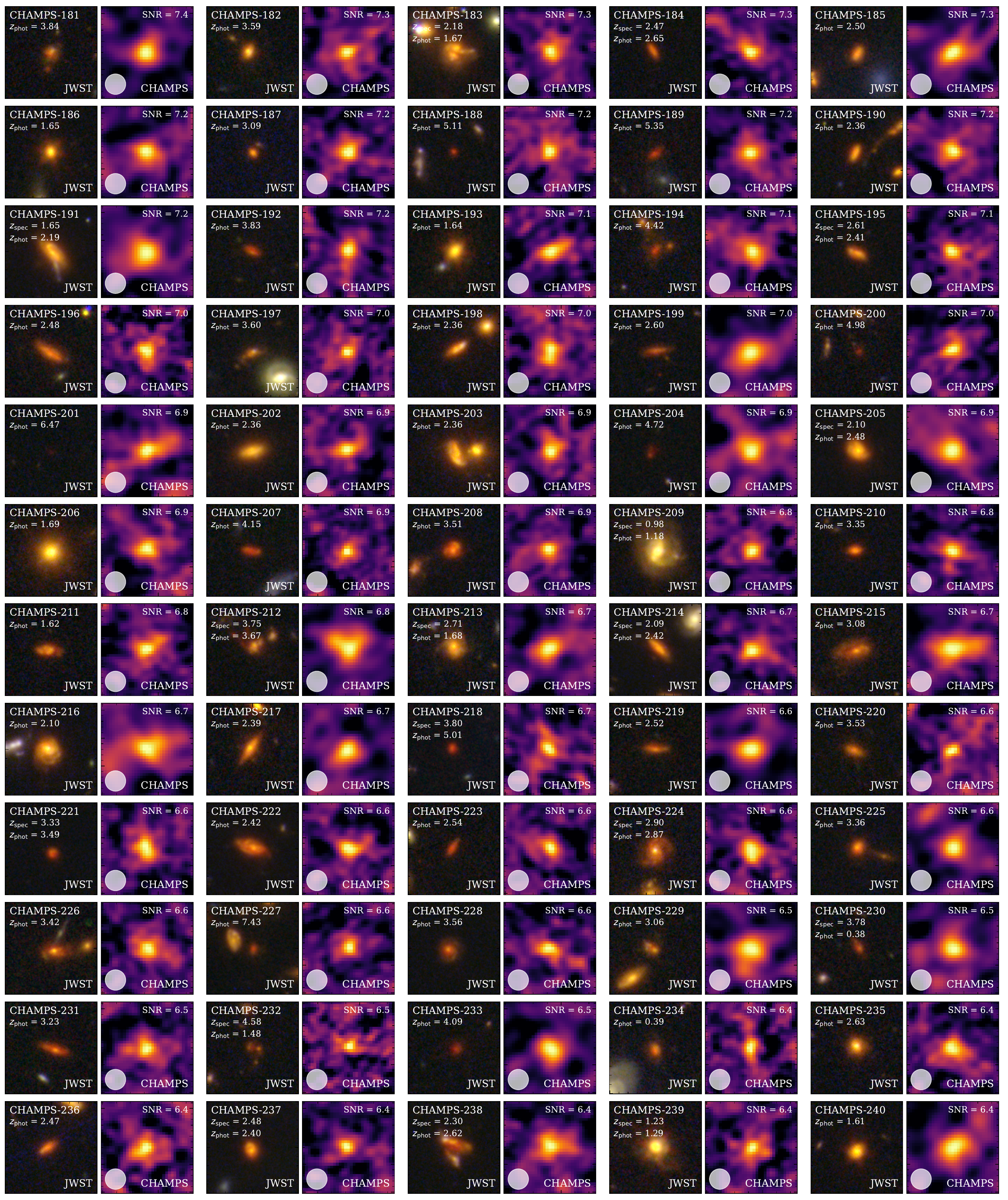}
\caption{Same as Figure~\ref{fig:collage1}.
\label{fig:collage4}}
\end{figure*}
%%%%%%%%%%%%%%%%%%%%%%%%%%%%%%%

%%% FIGURE: INDIVIDUAL GALAXIES %%%%
\begin{figure*}[b!]
\centering
\includegraphics[width=\textwidth,height=0.9\textheight,keepaspectratio]{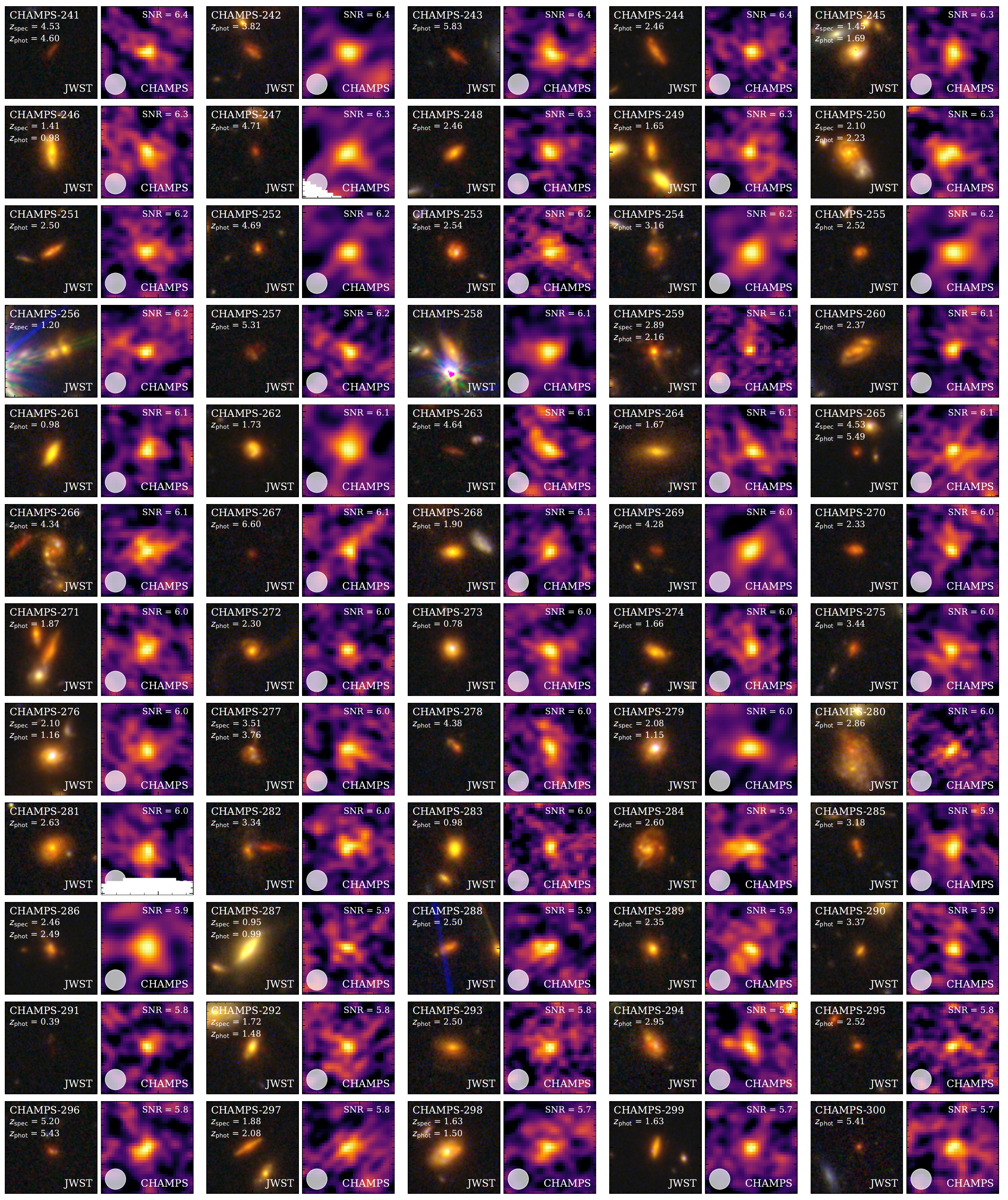}
\caption{Same as Figure~\ref{fig:collage1}.
\label{fig:collage5}}
\end{figure*}
%%%%%%%%%%%%%%%%%%%%%%%%%%%%%%%

%%% FIGURE: INDIVIDUAL GALAXIES %%%%
\begin{figure*}[b!]
\centering
\includegraphics[width=\textwidth,height=0.9\textheight,keepaspectratio]{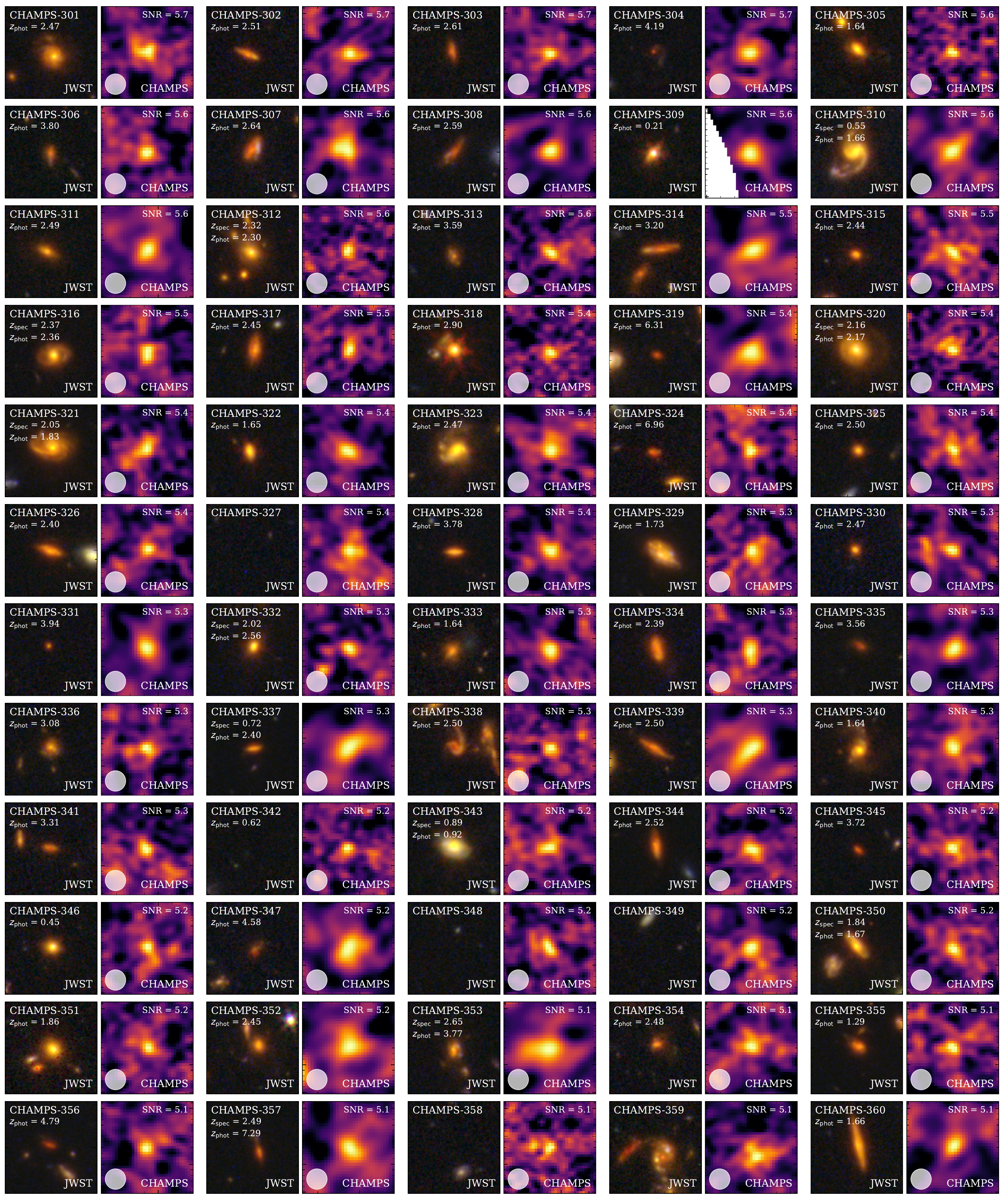}
\caption{Same as Figure~\ref{fig:collage1}. We note CHAMPS-327, CHAMPS-348, CHAMPS-349, and CHAMPS-358 are likely spurious sources as described in the text (see Section \ref{sec:blind_detections}).
\label{fig:collage6}}
\end{figure*}
%%%%%%%%%%%%%%%%%%%%%%%%%%%%%%%

%%% FIGURE: INDIVIDUAL GALAXIES %%%%
\begin{figure*}[t!]
\centering
\includegraphics[width=\textwidth,keepaspectratio]{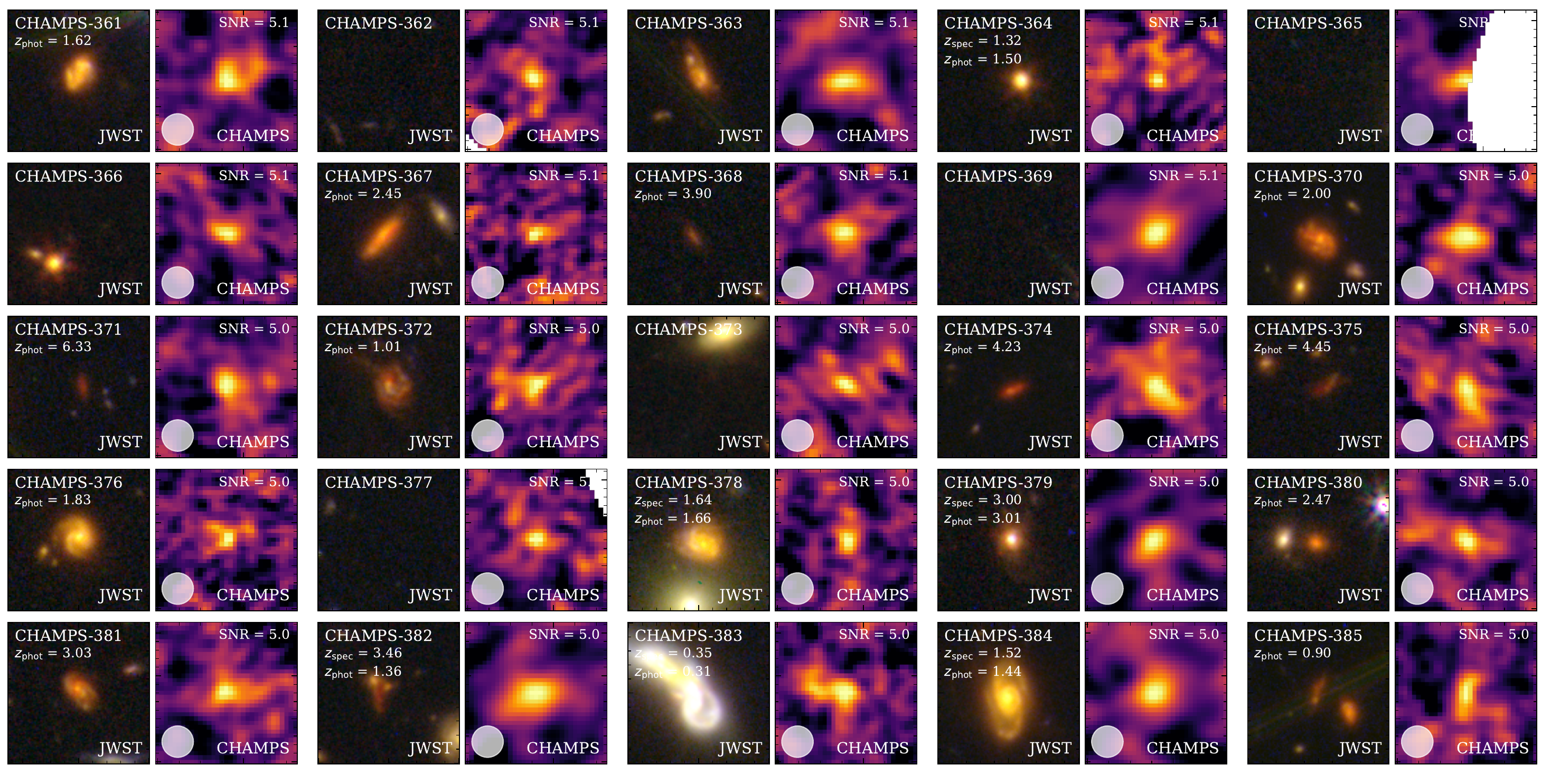}
\caption{Same as Figure~\ref{fig:collage1}. We note CHAMPS-362, CHAMPS-365, CHAMPS-369, CHAMPS-373, and CHAMPS-377 are likely spurious sources as described in the text (see Section \ref{sec:blind_detections}).
\label{fig:collage7}}
\end{figure*}
%%%%%%%%%%%%%%%%%%%%%%%%%%%%%%%
\label{fig:cutouts}

\clearpage

%%%%%%%%%%%%%%%%%%%%%%%%%%%%%%%
% %%% FIGURE: INDIVIDUAL GALAXIES %%%%
% \begin{figure*}[t]
% \centering
% \includegraphics[angle=0,width=\textwidth]{figures/collage-SnrGt5-7.pdf}
% \caption{Same as Figure~\ref{fig:collage1}. We note CHAMPS-362, CHAMPS-365, CHAMPS-369, CHAMPS-373, and CHAMPS-377 are likely spurious sources as described in the text.
% \label{fig:collage7}}
% \end{figure*}

\bibliography{references}{}

\allauthors
\end{document}